\documentclass[review]{elsarticle}
\usepackage[utf8]{inputenc}
\usepackage{amsmath}
\usepackage{graphicx}
\usepackage{float}
\usepackage{textcomp}
\usepackage{gensymb}
\usepackage{bm}
\usepackage{fullpage}
\usepackage[normalem]{ulem}
\begin{document}

\begin{frontmatter}

\title{Lattice Boltzmann Method for wave propagation in elastic solids with a
regular lattice: Theoretical analysis and validation}

\author[Polytechnique,M2P2]{Maxime Escande}
\author[jncasr]{Praveen Kumar Kolluru}
\author[upmc]{Louis Marie Cl\'eon}
\author[M2P2]{Pierre Sagaut\corref{corresAuthor}}
\cortext[corresAuthor]{Corresponding author}
\ead{pierre.sagaut@univ-amu.fr}

\address[Polytechnique]{Ecole Polytechnique, Route de Saclay, 91120 Palaiseau,
France.}
\address[M2P2]{Aix Marseille Univ, CNRS, Centrale Marseille, M2P2 UMR 7340,
13451 Marseille, France.}
\address[jncasr]{Engineering Mechanics Unit, Jawaharlal Nehru Centre for
Advanced Scientific Research, Jakkur, Bangalore 560064, India.}
\address[upmc]{Sorbonne Universit\'e, CNRS, Institut Jean Le Rond 
d'Alembert, F-75005 Paris, France.}

\begin{abstract}
The von Neumann stability analysis along with a Chapman-Enskog analysis is
proposed for a single-relaxation-time lattice Boltzmann Method (LBM) for wave 
propagation in isotropic linear elastic solids, using a regular D2Q9 lattice.
Different boundary conditions are considered: periodic, free surface, rigid
interface. An original absorbing layer model is proposed to prevent spurious 
wave reflection at domain boundaries. The present method is assessed considering
several test cases. First,  a spatial Gaussian force modulated in time by a 
Ricker wavelet is used as a source. Comparisons are made  with results obtained 
using a classical Fourier spectral method. Both P and S waves are shown to be 
very accurately predicted. The case of  Rayleigh surface waves is then addressed
to check the accuracy of the method.
\end{abstract}


\end{frontmatter}


\section{Introduction}

Lattice Boltzman Methods (LBM) are now a popular approach in Computational 
Fluid 
Dynamics (e.g. see \cite{kruger_book,guo_book}), with a wide range of 
applications, ranging from nearly-incompressible
flows to  multiphase flows, flows in porous media, combustion and recently to 
supersonic flows, e.g.  \cite{Jacob2018b,feng2019hybrid,Guo2020lbm}.
They exhibit a very high efficiency when compared to many classical  
Naviers-Stokes-based solvers, leading 
to a rapidly growing use for both academic 
and industrial purposes. Their main advantage comes from their explicit and 
compact character, along with the use of Cartesian grids. The common feature of 
Lattice-Boltzmann-like methods is that they rely on the following set of $m$ 
advection-collision equations:
\begin{equation}
\label{eq:LBM-generic}
\frac{\partial f_i}{\partial t} + \mathbf{c}_i \cdot \nabla f_i =
-\frac{1}{\tau} ( M[\mathbf{f}] - f_i ) , \quad i=0,m-1
\end{equation}
where the model used for the collision term in the right-hand side is the single
relaxation time Bhatnagar-Gross-Krook(BGK) model. Here, $\mathbf{c}_i$ is a 
set of arbitrary solution-independent velocities, $M[\mathbf{f}]$ is a function 
of $f$ designed to recover the targeted macroscopic equations (the Maxwellian 
in classical hydrodynamics) and $\tau$   a relaxation time. The computational 
efficiency of LBM-like  approaches comes from the Strang splitting used to 
decouple the advection and the collision step \cite{dellar2014}: the advection 
step simplifies as a swap in memory without additional floating point 
operations, while the collision step is strictly local in each cell. The 
macroscopic quantities such as density, velocity and pressure in fluid 
mechanics 
are recovered by post-processing the functions $f_i$. A key element in LBM is 
the choice of the lattice, i.e. the set of $m$ nodes (one per couple $(f_i, 
\mathbf{c}_i)$) needed to solve Eq. \eqref{eq:LBM-generic} at a given node. The 
dimension $N$ comes from a trade-off between the accuracy, 
 stability and computational efficiency of the method. A lattice is generally 
denoted $DnQm$, where $n$ and $m$ are the space dimension and the number of 
nodes in the lattice, respectively.

While LBM originally appeared as a way to solve numerically the Boltzmann
equation that statistically describes fluid flow physics, it can also be 
interpreted and extended as a smart way to solve sets of PDEs, especially for 
wave propagation \cite{guangwu2000}. In that case, 
they must be seen as a purely numerical way to solve an equation, starting from 
an ad hoc change of variables to recover a set of coupled advection-reaction 
equations. In order to preserve the efficiency of the method, the advection 
operator must be linear and the reaction term should be written as a relaxation 
term. Using this approach, LBM-like methods have been proposed for many 
physical 
models, e.g. the heat conduction equation (e.g. \cite{jiaung2001}), the 
Schr\"odinger equation (e.g. \cite{zhong2006})  and Maxwell's equations (e.g. 
\cite{hasanoge2011,liu2014}).

The present paper addresses the issue of 
developing a LBM-like method for the 
propagation of waves in elastic solids. Up to now, extension of LBM for solid 
mechanics has been addressed by a very few authors only, and a general fully 
satisfactory method is still to be developed.

As a matter of fact, a LBM approach has been proposed to solve the Lam\'e
equation, which governs the static deformation for linear elastic solids
\cite{yin_direct_2016}. An unsteady model, that was able to simulate wave 
propagation in Poisson solids has been proposed in 
\cite{obrien_lattice_2012}, using a finite difference propagation operator with 
a  limiter to enhance the 
stability of the method. This approach is restricted to the case of Poisson 
solids, and necessitates the use of very small time steps to prevent numerical 
instability. A more general method was recently proposed by 
\cite{jmurthy_lattice_2018}, which is not restricted to Poisson solids and it 
is 
based on the collide-stream implementation that characterizes the fluid LBM 
approaches. This method is based on the use of complex crystallographic 
lattices. The case of propagation of shock waves in solids was addressed in the 
1D case in \cite{xiao2007}, in which a finite element method is used for the 
propagation step.

Previous unsteady methods use what is referred to as "multi-speed" lattices,
since the lattices involve a large number of nodes resulting in wide 
computational stencils. In some of them,  lattice have either half-speeds: 
velocities in between the neighbour lattice site (so a double staggered grid is 
needed), or double-speeds (or more), i.e. velocities above the neighbour 
lattice 
site. O'Brien et al. \cite{obrien_lattice_2012} uses 
 a D2Q13 (D2Q9 + $(\pm 2,0)$ form vectors) in 2D or a D3Q25 (D3Q19 + $(\pm 
2,0,0)$) in 3D. Murthy et 
al. \cite{jmurthy_lattice_2018} uses crystallographic lattices 
\cite{namburi_crystallographic_2016}: RD3Q27 (a 
D3Q27 including half-speeds) and 
RD3Q41 \cite{kolluru2020lattice}: (D3Q27 + 8 half-speeds + 6 double-speeds). 
Such sets have a higher-order truncation, so 
they are sometimes necessary to recover the good macroscopic behaviour. However, 
for complex geometries dealing with many interfaces with corners or non-planar 
boundaries, multi-speed lattices may be cumbersome to implement.

The case of viscoelastic fluid flows has been addressed by several authors,
e.g. \cite{ispolatov2002, lallemand2003, malaspinas2010, phillips2011,
frantziskonis_lattice_2011, dellar2014,gupta2015}, leading to the definition of 
several Lattice Boltzmann models for this type of rheology. A common difference 
between all these models and those sought for the case of elastic solids is that
the former all work on the fluid velocity due to the presence of a flow, while 
the displacement is the natural physical quantity in the solid case. Even 
though a viscoelastic solid behavior for wave propagation can be recovered as 
an asymptotic case of a general viscoelastic fluid, the two families of 
numerical methods will exhibit different properties. The possibility to capture 
P and S waves in a viscoelastic fluid by adding the adequate stresses as a body 
force to a classical fluid LBM was demonstrated in 
\cite{frantziskonis_lattice_2011} via 2D numerical experiments.

It is worth noting that Lattice Boltzmann Methods share some features of the
Elastic Lattice Models  (ELM) used to study wave propagation in solids, e.g. 
see \cite{obrien2004, obrien2011, obrien2014, obrien2009, vallegarcia2003, 
xia2017}. It is reminded that in ELM, nodes in the lattice are coupled via 
linear and angular springs. The main common point is the use of a discrete 
lattice of neighbouring nodes to advance the solution in time at a given node.
The second one is that the collision model in LBM and the spring model in ELM 
are tuned to recover the targeted macroscopic continuous elastic wave equations 
via ad hoc multiscale expansion. In most cases, elastic energy conservation and 
Taylor expansion are used in ELM (e.g. \cite{polyzos2012}), while collisional 
invariant preservation and Chapman-Enskog expansion are used in LBM. But some 
important differences must be noticed. First, ELM works directly on macroscopic 
quantities (acceleration, velocity) while LBM works on mesoscopic quantities 
$f_i$. Second, ELM is based on Newton's second law, each nodes in the lattice 
being tied to the node under consideration by a spring, the characteristic of 
the later being tuned to recover the desired effect. In LBM, one solves coupled 
advection-relaxation equations, leading to different numerical features.

Therefore, defining a LBM method based on a  regular lattice is a first step
towards simulation of complex solid media. It is proposed here to adapt Murthy's
method to regular lattices (see Section \ref{sec:lbm-solid-model}), and to 
analyze its features by carrying  a Linear Stability Analysis (see Section 
\ref{sec:LBM-LSA})  and deriving rigorously the associated macroscopic equations
via a multiscale analysis that mimics the Chapman-Enskog expansion used in the 
fluid case (see Section \ref{sec:CE-solid}). The method is then validated 
considering propagation of S and P waves in an infinite medium (see Section 
\ref{sec:SPwaves}).

Another extension of the model proposed in \cite{jmurthy_lattice_2018} is then
performed by addressing finite computational domains and boundary conditions 
(see Section \ref{sec:BCs}) , along with the definition of an original sponge 
layer technique to avoid spurious wave reflection. These developments are 
validated considering the propagation of Rayleigh surface waves.

\section{Lattice Boltzmann Method for wave propagation in elastic solids}
\label{sec:lbm-solid-model}
The LB method presented here is based on the one proposed by Murthy et al. 
\cite{jmurthy_lattice_2018} which  employs higher-order Crystallographic 
Lattices \cite{namburi_crystallographic_2016, kolluru2020lattice}. These 
lattices were shown to result in a gain in terms of compactness of the 
computational stencil and number of unknowns per node. The proposed LB model
for elastic solids shares some features with the well-known basic LBM 
for fluids, more precisely with weakly compressible isothermal LBM with BGK 
collision operator. In this work, we extend this LB model for regular 
lattices, particularly to the widely used D2Q9 lattice instead of a 
crystallographic lattice.

The goal here is to recover the Navier equation for homogeneous linear isotropic
elastic solids, which yields the following  wave propagation equation:
\begin{equation}
\rho_0 \partial_t^2 u_\alpha = (\lambda+\mu)\partial_\alpha \partial_\beta
u_\beta + \mu \partial_\beta^2 u_\alpha +F_\alpha
\label{NavierU}
\end{equation}
where $\rho_0$ is the density (assumed to be  homogeneous), $u_\alpha$ is the
displacement, $\lambda$ and $\mu$ are the Lam\'e coefficients and $F_\alpha$ is 
a bulk external force. Introducing the mass flux $j_\alpha=\rho \partial_t 
u_\alpha$, we have 
\begin{equation}
\partial_t^2 j_\alpha = \frac{\lambda+\mu}{\rho_0}\partial_\alpha \partial_\beta
j_\beta + \frac{\mu}{\rho_0} \partial_\beta^2 j_\alpha + \frac{1}{\rho_0} 
\partial_t F_\alpha
\label{NavierJ}
\end{equation}

By separating a longitudinal and transverse parts in this equation, one can
recover both P waves (compressive) and S waves (shear) with their respective 
wave speeds being
\begin{align}
\begin{split}
v_P=\sqrt{\frac{\lambda+2\mu}{\rho_0}}, \quad v_S=\sqrt{\frac{\mu}{\rho_0}}.
\end{split}
\label{VpVs}
\end{align}
The velocity ratio of the primary and secondary waves is
\begin{equation}
    \frac{v_P}{v_S}=\sqrt{\frac{2-2\nu}{1-2\nu}},
    \label{VpVsRatio}
\end{equation}
with Poisson ratio $\nu$ defined as
\begin{equation}
    \nu=\frac{\lambda}{2(\lambda+\mu)}.
    \label{nu}
\end{equation}

Equation \eqref{NavierJ} can be recast  with a moment chain with a source term
as follows \cite{godunov2003}
\begin{align}
 \begin{split}
&\partial_{t}\rho + \partial_{\alpha}j_{\alpha} = 0\\
&  \partial_{t}j_{\alpha} + \partial_{\beta}P_{\alpha\beta} = \frac{\mu-\lambda}
{\rho_0} \partial_\alpha \rho + F_\alpha\\
&  \partial_{t}P_{\alpha\beta} + \partial_\gamma Q_{\alpha\beta\gamma}^{eq}=0
 \end{split}
 \label{MomentChain}
\end{align}
with 
\begin{equation}
    Q_{\alpha\beta\gamma}^{eq}
    =\frac{\mu}{\rho_0}\left(j_\alpha\delta_{\beta\gamma}+
    j_\beta\delta_{\alpha\gamma}+
    j_\gamma\delta_{\alpha\beta}\right)
    \label{Q}
\end{equation}
and $P_{\alpha\beta}$ behaving as the stress tensor but with the opposite sign,
i.e. $P_{\alpha\beta}=-\sigma_{\alpha\beta}$, leading to a set of conservation 
equations whose mathematical structure is close to the one used for 
non-Newtonian fluids, with the important difference that the mass flux appears 
in place of the momentum.

 The  force term in the mass flux equation is necessary for non-Poisson solids
 (Poisson solids are solids with  $\lambda=\mu$, i.e. $\nu=0.25$), as noticed 
by \cite{jmurthy_lattice_2018} when extending the model for Poisson solids 
proposed by O'Brien \cite{obrien_lattice_2012}. We see through the conservation 
of mass that the density is not taken constant and homogeneous but is considered
as an independent variable, while $\rho_0$ is still used as a constant.
From now on, we  merge the elastic force $\frac{\mu-\lambda}{\rho_0}
\partial_\alpha \rho$ and external force $F_\alpha$ in a 
generic source term
\begin{equation}
S_\alpha=\frac{\mu-\lambda}{\rho_0}\partial_\alpha \rho + F_\alpha
\label{eq:S}
\end{equation}

In practice, the spatial derivative $\partial_\alpha \rho$ is computed via a 
second-order centered finite difference scheme
\begin{equation}
    \partial_\alpha \rho(\pmb{x})=
    \frac{\rho(\pmb{x}+\pmb{\Delta x_\alpha})
    -\rho(\pmb{x}-\pmb{\Delta x_\alpha})}
    {2\Delta x_\alpha}.
    \label{FDrho}
\end{equation}

We define the forcing term $S_i$ as
\begin{equation}
S_i=w_i \frac{c_{i\alpha} S_\alpha}{b^2}
\label{Si}
\end{equation}

The macroscopic quantities are recovered computing the moments of $f_i^{eq}$
according to the following relations
\begin{align}
\begin{split}
    &\sum_i f_i^{eq}=\rho\\
    &\sum_i f_i^{eq}c_{i\alpha}=j_\alpha
    -\frac{\Delta t}{2}S_\alpha\\
    &\sum_i f_i^{eq}c_{i\alpha}c_{i\beta}=P_{\alpha\beta}\\
    &\sum_i f_i^{eq}c_{i\alpha}c_{i\beta}c_{i\gamma}
    =Q_{\alpha\beta\gamma}^{eq}
    =b^2\left(j_\alpha\delta_{\beta\gamma}+
    j_\beta\delta_{\alpha\gamma}+
    j_\gamma\delta_{\alpha\beta}\right)\\
\end{split}
\label{MomentsFeq}
\end{align}
with $b$ being the ``sound speed'' of the set, matching the S wave speed 
$v_S^2=\mu/\rho_0$ in Eq. \eqref{Q}. So one can see that the classical 
concept of sound speed in LBM for fluids $c_s$ matches with the S wave 
speed here. So for regular lattices, one obtains $v_S = b =1/\sqrt{3}$.

So far, zeroth-, first- and third-order moments in \eqref{MomentsFeq} are
consistent with those defined in fluid LBM:
\begin{align}
    \begin{split}
        &\sum_i f_i^{eq,fluids}=\rho\\
        &\sum_i f_i^{eq,fluids}c_{i\alpha}=\rho v_\alpha\\
        &\sum_i f_i^{eq,fluids}c_{i\alpha}c_{i\beta}c_{i\gamma}
        =\rho c_s^2\left(v_\alpha\delta_{\beta\gamma}+
        v_\beta\delta_{\alpha\gamma}+
        v_\gamma\delta_{\alpha\beta}\right)\\
    \end{split}
    \label{MomentsFeqFluids}
\end{align}
where $v_\alpha=\partial_t u_\alpha=j_\alpha/\rho$ is the fluid velocity.\\

The equilibrium distribution functions for the elastic solid LBM are given by
\begin{equation}
\label{eq:equilibrium-solid}
f_i^{eq}=w_i\left(\rho+\frac{j_\alpha c_{i\alpha}}{b^2}+    
\frac{P_{\alpha\beta}^n\left(c_{i\alpha}c_{i\beta}-b^2 
\delta_{\alpha\beta}\right)}{2 b^4}\right)
\end{equation}
with
\begin{equation}
    P_{\alpha\beta}^n=P_{\alpha\beta}-\rho b^2\delta_{\alpha\beta}
    \label{Pn}
\end{equation}

This is the same form as the equilibrium distribution for fluids,  replacing
$u_\alpha u_\beta$ by $P_{\alpha\beta}^n$:
\begin{equation}
f_i^{eq,fluids}=w_i\rho\left(1+\frac{u_\alpha c_{i\alpha}}{b^2}+
    \frac{u_\alpha u_\beta \left(c_{i\alpha}c_{i\beta}-b^2 \delta_{\alpha\beta}
    \right)}{2 b^4}\right)
\end{equation}

This is consistent with the second order moment in equation \eqref{MomentsFeq}
since for fluids
\begin{equation}
\sum_i f_i^{eq,fluids}c_{i\alpha}c_{i\beta}=\rho u_\alpha u_\beta + \rho b^2 
    \delta_{\alpha\beta} 
    \equiv P_{\alpha\beta}
    \label{MomentsFeqFluids2}
\end{equation}

The resulting discrete Lattice Boltzmann equation with the source terms is: 
\begin{equation}
    f_i\left(\pmb{x}+\pmb{c_i}\Delta t,t+\Delta t\right)
    -f_i\left(\pmb{x},t\right)=
    -\frac{\Delta t}{\tau}\left(f_i\left(\pmb{x},t\right)-
    f_i^{eq}\left(\pmb{x},t\right)\right)+
    \Delta t \left(1-\frac{\Delta t}{2\tau}\right)S_i
    \label{LBGK}
\end{equation}

A second order accurate explicit time integration can be obtained using the same
change of variables as in fluid LBM. In the rest of the article we consider a 
non-dimensionalized system, with the simplest lattice units $\Delta t=1$, 
$\Delta x_\alpha=1$.\\

In the next section, we present a Chapman-Enskog analysis for this model.
\section{Associated macroscopic conservation equation: multiscale 
Chapman-Enskog-type analysis}
\label{sec:CE-solid}

As in the fluid LBM case, it is assumed that the solution $f_i$ corresponds to 
a small perturbation about the equilibrium state $f_i^{eq}$.
To recover the macroscopic evolution equations associated to Eqs. \eqref{LBGK}, 
\eqref{eq:equilibrium-solid} and \eqref{eq:S}, it is proposed to mimic the 
Chapman-Enskog multiscale expansion classically used in fluid LBM to this end.
Here, this procedure is purely mathematical, since there is no link with the 
physical Boltzmann equation for statistical fluid physics in the case of solid 
LBM. Introducing the small parameter $\epsilon$ (which is the Knudsen number in 
the dilute gas case), one considers the following expansion

\begin{equation}
    f_i=f_i^{eq}+\epsilon f_i^{(1)}+\epsilon^2 f_i^{(2)}+...
    \label{expansion}
\end{equation}
As in the classical Chapman-Enskog expansion, it is assumed that both $S_i$ and
$S_\alpha$ are of order $O(\epsilon)$
\begin{equation}
S_\alpha=\epsilon S_\alpha^{(1)}
\label{Sepsilon}
\end{equation}
and time and spatial derivatives are expanded as
\begin{align}
    \begin{split}
        &\partial_t=\epsilon \partial_t^{(1)}+
        \epsilon^2 \partial_t^{(2)}+...\\
        &\partial_\alpha=\epsilon \partial_\alpha^{(1)}.
    \end{split}
\end{align}

From the mass, momentum and stress conservation equations 
Eq. \eqref{MomentChain} we obtain the following solvability conditions. It 
should be noted that the forcing is responsible for the non-zero terms:
\begin{align}
\begin{split}
    &\sum_i f_i^{neq}=-\frac{\Delta t}{2}\sum_i S_i^{(1)}=0,\\
    &\sum_i f_i^{neq}c_{i\alpha}=-\frac{\Delta t}{2}\sum_i S_i^{(1)} 
    c_{i\alpha}=-
    \frac{\Delta t}{2}S_\alpha^{(1)},\\
    &\sum_i f_i^{neq}c_{i\alpha}c_{i\beta}=-\frac{\Delta t}{2}\sum_i S_i^{(1)} 
    c_{i\alpha}c_{i\beta}=0
\end{split}
\label{solvability1}
\end{align}
Since $S_\alpha^{(1)}\sim O(\epsilon)$ it only affects $f_i^{(1)}$, 
hence, the order-by-order solvability conditions are
\begin{align}
\begin{split}
    &\sum_i f_i^{(n\geq1)}=0,\\
    &\sum_i f_i^{(1)}c_{i\alpha}=-\frac{\Delta t}{2}S_\alpha^{(1)},\text{ and } 
    \sum_i f_i^{(n\geq2)}c_{i\alpha}=0,\\
    &\sum_i f_i^{(n\geq1)}c_{i\alpha}c_{i\beta}=0
\end{split}
\label{solvability2}
\end{align}
Incompressible hydrodynamic equations (for Newtonian fluids) lack the 
last condition as only mass and momentum are conserved quantities.

By using Eq. \eqref{expansion} in a Taylor expansion of the Lattice 
Boltzmann equation Eq. \eqref{LBGK}, and by identifying the terms of order 
$O(\epsilon)$ and $O(\epsilon^2)$,  we have:
\begin{align}
\begin{split}
\left(\partial_t^{(1)}+c_{i\alpha}\partial_\alpha^{(1)}\right)f_i^{eq}-
\left(1-\frac{\Delta t}{2\tau}\right)w_i \frac{c_{i\beta}S_\beta^{(1)}}{b^2}&=
-\frac{1}{\tau}f_i^{(1)}\\
\partial_t^{(2)}f_i^{eq}+\left(1-\frac{\Delta t}{2\tau}\right)
\left(\partial_t^{(1)}+c_{i\alpha}\partial_\alpha^{(1)}\right)
\left(f_i^{(1)}+
\frac{\Delta t}{2}w_i \frac{c_{i\beta} S_\beta^{(1)}}{b^2}\right)&=
-\frac{1}{\tau}f_i^{(2)}\\
\end{split}
\label{CE}
\end{align}

Taking the zeroth through third moment at $O(\epsilon)$ from Eq. \eqref{CE}, 
one obtains
\begin{align}
\begin{split}
&\partial_t^{(1)}\rho+\partial_\alpha^{(1)}j_\alpha=0\\
&\partial_t^{(1)}j_\alpha+\partial_\beta^{(1)}P_{\alpha\beta}=S_\alpha^{(1)}\\
&\partial_t^{(1)}P_{\alpha\beta}+\partial_\gamma^{(1)}Q_{\alpha\beta\gamma}^{eq}
=0\\
&\partial_t^{(1)}Q_{\alpha\beta\gamma}^{eq}+\partial_\kappa^{(1)}
R_{\alpha\beta\gamma\kappa}^{eq}=-\frac{1}{\tau}Q_{\alpha\beta\gamma}^{(1)}+
b^2\left(1-\frac{\Delta t}{2\tau}\right)
\left(S^{(1)}_\alpha\delta_{\beta\gamma}+
S^{(1)}_\beta \delta_{\alpha\gamma}+
S^{(1)}_\gamma \delta_{\alpha\beta} \right),
\end{split}
\label{CEmoments1}
\end{align}
at  $O(\epsilon^2)$ we have
\begin{align}
\begin{split}
&\partial_t^{(2)}\rho=0\\
&\partial_t^{(2)}j_\alpha=0\\
&\partial_t^{(2)}P_{\alpha\beta}+\left(1-\frac{\Delta t}{2\tau}\right)
\partial_\gamma^{(1)}Q_{\alpha\beta\gamma}^{(1)}=
-b^2\frac{\Delta t}{2}\left(1-\frac{\Delta t}{2\tau}\right)\left(
\partial_\gamma^{(1)}S_\gamma^{(1)}\delta_{\alpha\beta}+
\partial_\alpha^{(1)}S_\beta^{(1)}+
\partial_\beta^{(1)}S_\alpha^{(1)}\right).
\end{split}
\label{CEmoments2}
\end{align}

As can be seen from the above set of equations the additional term 
$Q_{\alpha\beta\gamma}^{(1)}$ appearing in Eq. \eqref{CEmoments2} depends on 
$R_{\alpha\beta\gamma\kappa}^{eq}$ via Eq. \eqref{CEmoments1} which in turn 
depends on the choice of the discrete velocity model.
For regular lattices,  \emph{D2Q9, D3Q15, D3Q19 and D3Q27},  satisfying the 
following isotropy conditions
\begin{align}
\begin{split}
&\sum_i w_i=1,\quad
\sum_i w_i c_{i\alpha}c_{i\beta}=b^2 \delta_{\alpha\beta},\\
&\sum_i w_i c_{i\alpha}c_{i\beta}c_{i\gamma}c_{i\kappa}
=b^4 \Delta^{(4)}_{\alpha\beta\gamma\kappa}=b^4\left(\delta_{\alpha\beta}
\delta_{\gamma\kappa}+\delta_{\alpha\gamma}\delta_{\beta\kappa}+
\delta_{\alpha\kappa}\delta_{\beta\gamma}\right),\\
&\sum_i w_i c_{i\alpha}c_{i\beta}c_{i\gamma}c_{i\kappa}c_{i\theta}c_{i\psi}=
b^6 \left(\Delta^{(6)}_{\alpha\beta\gamma\kappa\theta\psi} 
- 6\delta_{\alpha\beta\gamma\kappa\theta\psi}\right)\\
&=b^6\left(
\delta_{\alpha\beta}\Delta^{(4)}_{\gamma\kappa\theta\psi}+
\delta_{\alpha\gamma}\Delta^{(4)}_{\beta\kappa\theta\psi}+
\delta_{\alpha\kappa}\Delta^{(4)}_{\beta\gamma\theta\psi}+
\delta_{\alpha\theta}\Delta^{(4)}_{\beta\gamma\kappa\psi}+
\delta_{\alpha\psi}\Delta^{(4)}_{\beta\gamma\kappa\theta}
-6\delta_{\alpha\beta\gamma\kappa\theta\psi}\right),\\
\end{split}
\label{isotropy}
\end{align}
the modified chain $\epsilon\eqref{CEmoments1}+ \epsilon^2\eqref{CEmoments2}$ 
reduces  to
\begin{align}
\begin{split}
&\partial_{t}\rho + \partial_{\alpha}j_{\alpha} = 0\\
&\partial_{t}j_{\alpha} + \partial_{\beta}P_{\alpha\beta} = S_\alpha\\
&\partial_t P_{\alpha\beta} + \partial_\gamma Q_{\alpha\beta\gamma}^{eq}
=\left(\tau-\frac{\Delta t}{2}\right)b^2\partial_\gamma\left(\partial_\alpha 
P^n_{\beta\gamma}+
\partial_\beta P^n_{\alpha\gamma}+
\partial_\gamma P^n_{\alpha\beta}
-3\partial_\gamma P_{\kappa\kappa}^n\delta_{\alpha\beta\gamma}\right)
\end{split}
 \label{MomentChain2}
\end{align}
The details of the calculations are given in Appendix A for the sake of brevity.

It should be noted that the last relation in Eq. \eqref{isotropy} is specific 
to regular lattices and takes into account the lack of isotropy of such 
sets. Ideally, the correct 6th order moment should be 
$\Delta^{(6)}_{\alpha\beta\gamma\kappa\theta\psi}$, but is known to be 
satisfied only by very high-order on-lattice models 
\cite{chen_fundamental_2008}.

{The multiscale analysis reveals the existence of the source term in the 
associated macroscopic evolution equation for the stresses. This source term is 
of diffusive/dissipative nature, and therefore will tend to damp and smooth the 
stresses $P^n_{\alpha\beta}$. It is proportional to $b^2\left(\tau-\Delta 
t/2\right)$, and therefore it can be tuned by adjusting the time step $\Delta 
t$.}

\section{Linear Stability Analysis}
\label{sec:LBM-LSA}
To study the stability of the scheme, the modified moment chain 
\eqref{MomentChain2} could be used to find the modified Navier equation and 
study its modes with plane analysis. However, finding the modified Navier 
equation  doesn't carry all the useful information on the LBM method.

In order to supplement the modified equation analysis, the von Neumann analysis 
of the distribution functions $f_i$ is now carried out (see \cite{sengupta2007} 
for a discussion of the correct spectral analysis and 
\cite{wissocq_extended_2019} for the interpretation within the LBM framework). 
We consider disturbances of the plane wave form:

\begin{equation}
    f_i(\pmb{x},t)=f_i^0 \exp(-i(\omega t+\pmb{k\cdot x}))
    \label{fiVN}
\end{equation}

For the vector $F=\left(f_0,f_1,...,f_{q-1}\right)^T$, a matrix equation is 
obtained by using the moments \eqref{MomentsFeq} of  Eq. \eqref{LBGK}
\begin{equation}
    F(t+\Delta t)=M\cdot F(t)
    \label{Fmatrix}
\end{equation}

where the coefficients $m_{ij}$ of $M$ are the following
\begin{align}
 \begin{split}
 m_{lj}=\exp(-ik_\gamma c_{l\gamma}\Delta t)
 \left[\left(1-\frac{\Delta t}{\tau}\right) \delta_{lj}
 +w_l\frac{\Delta t}{\tau}\left( 1 + \frac{c_{l\alpha}c_{j\alpha}}{b^2}
 +i\Delta t\Lambda\frac{c_{l\alpha} \sin(k_\alpha \Delta x_\alpha)}{2 
 \Delta x_\alpha}\right.\right.\\
\left.\left.+\frac{1}{2b^4}\left(c_{l\alpha}c_{l\beta}-b^2\delta_{\alpha\beta}  
\right)\left(c_{j\alpha}c_{j\beta}-b^2\delta_{\alpha\beta}
\right)\right) +\left(1-\frac{\Delta t}{2\tau}\right)\Delta t \Lambda b^2
\frac{c_{l\alpha} \sin(k_\alpha \Delta x_\alpha)}{\Delta x_\alpha}\right]
\end{split}
\label{VN}
\end{align}
with
\begin{equation}
\Lambda=\frac{\mu-\lambda}{\rho_0 b^2}=\frac{1-4\nu}{1-2\nu}
\end{equation}
using $\mu/\rho_0=b^2$ and Eq. \eqref{nu}. 
In Eq. \eqref{VN}, the matrix index $i$ is replaced by $l$ to avoid any 
confusion with the imaginary unit $i^2=-1$.

\begin{figure}
\centering
\includegraphics[scale=0.35]{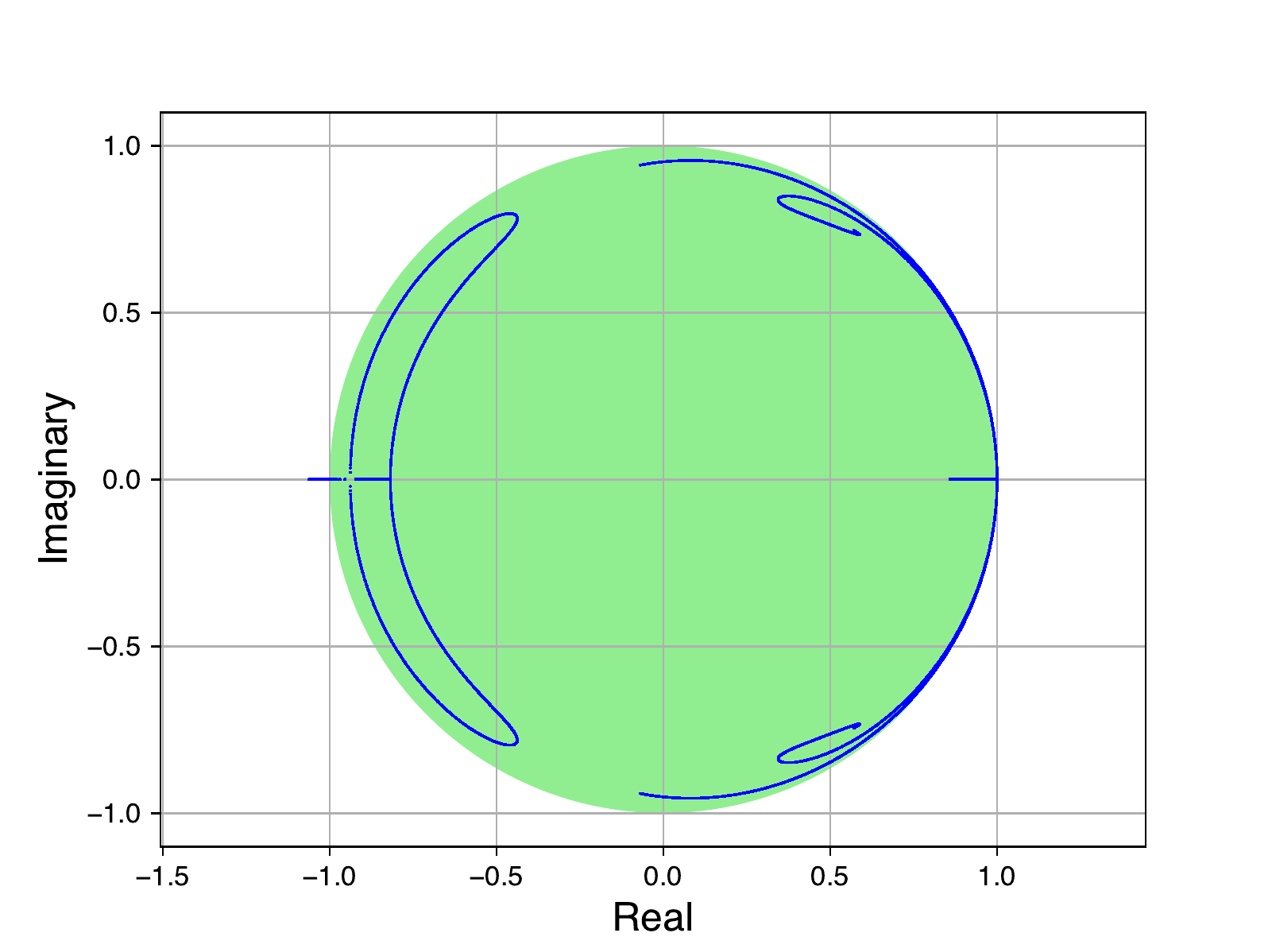}
\includegraphics[scale=0.35]{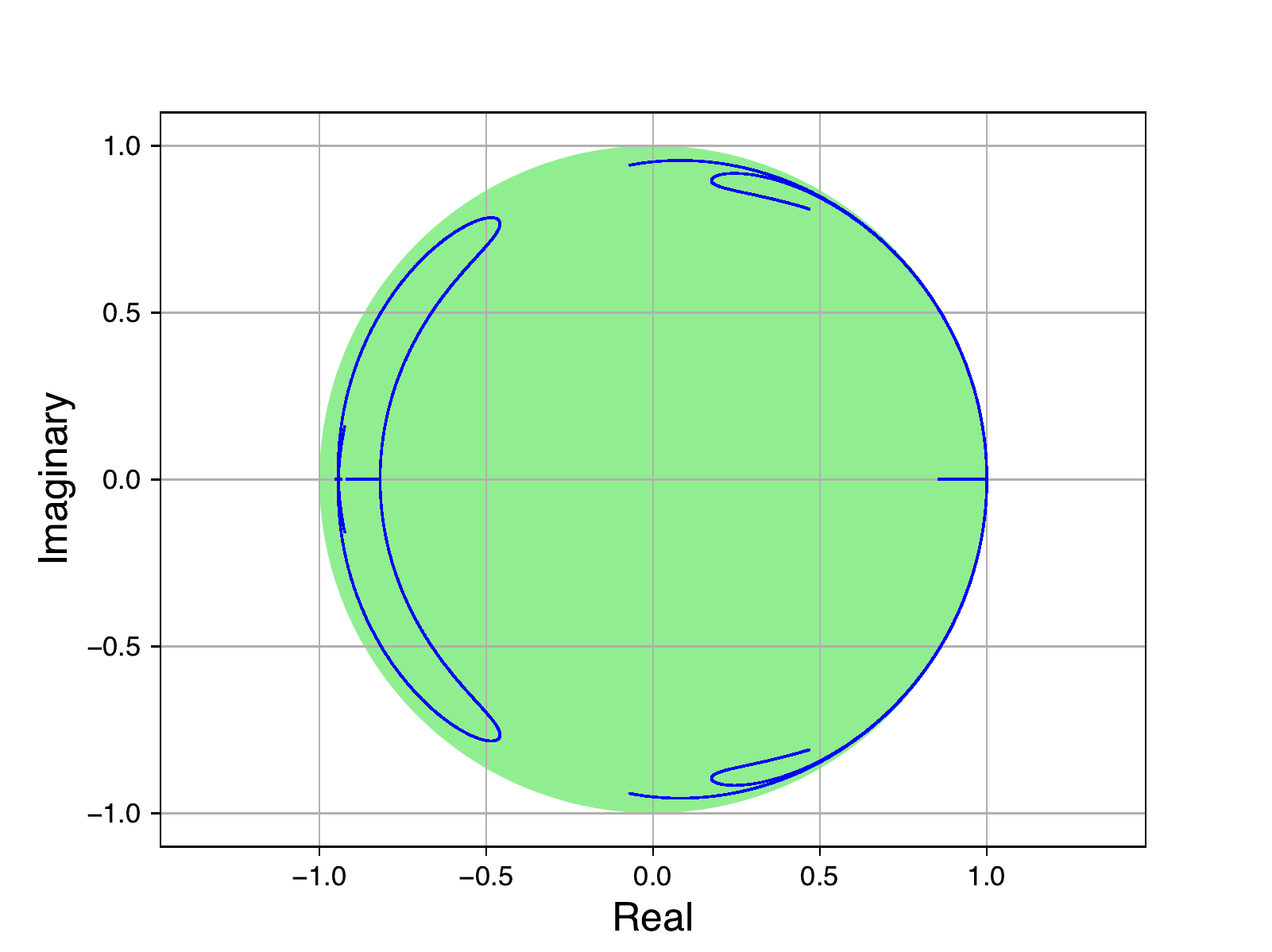}
\includegraphics[scale=0.35]{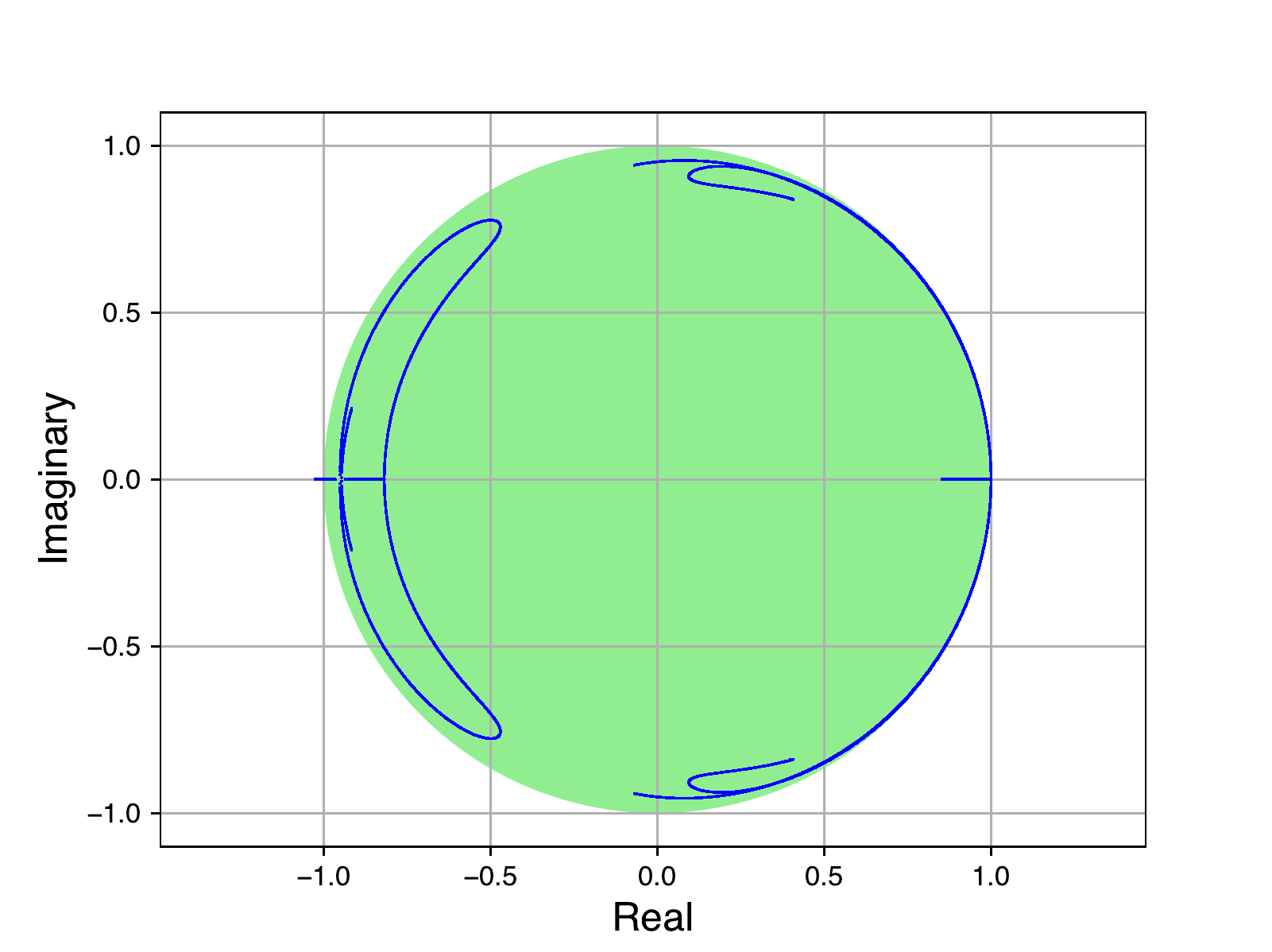}
\includegraphics[scale=0.35]{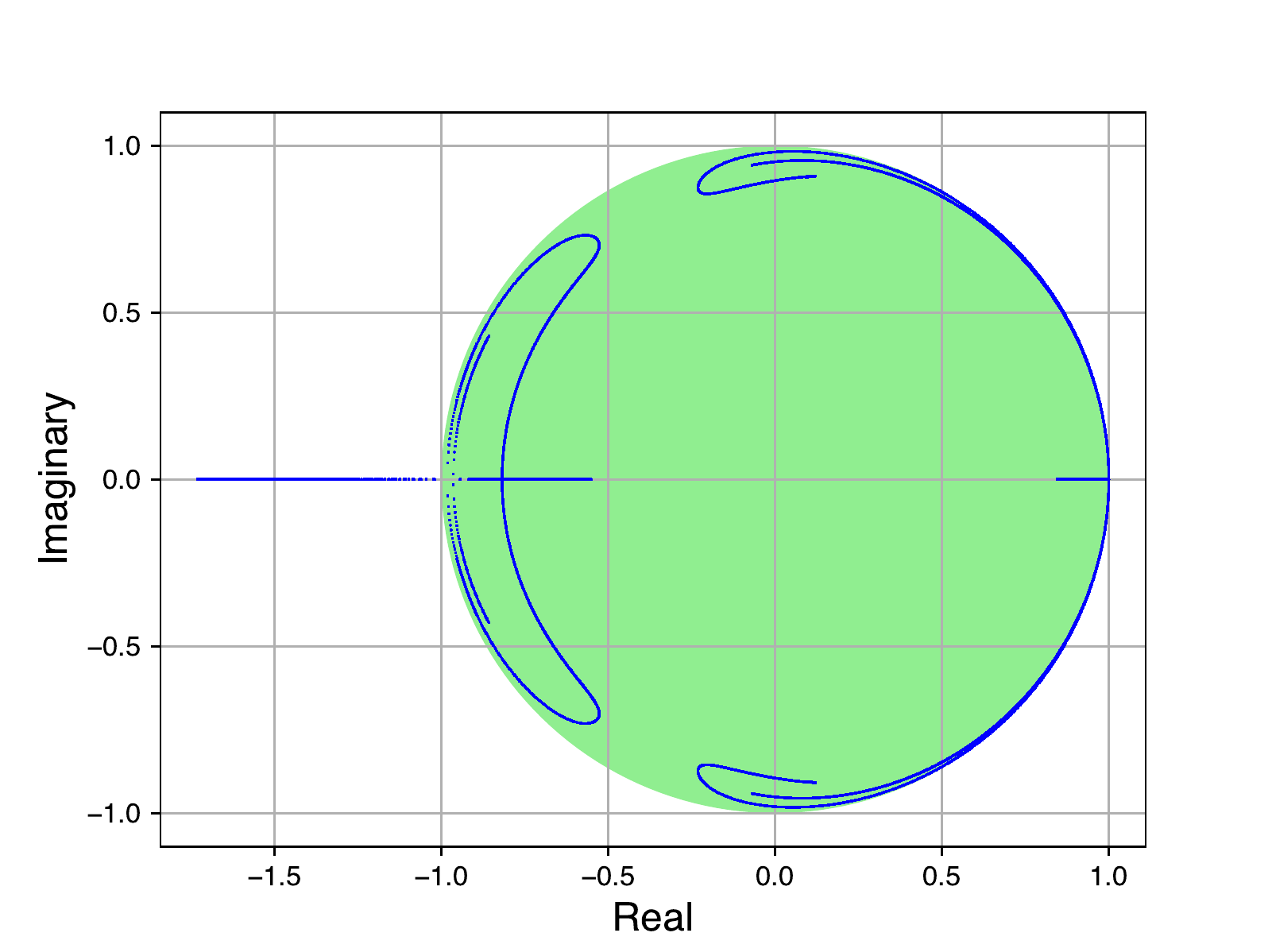}
\caption{Eigenvalues and stability disks for different Poisson ratio. Top-left: 
0; Top-right: 0.25; Bottom-left:0.3; Bottom-right: 0.4.
The stability disk is plotted in green, and the eigenvalues in blue (refer to 
the web version of the article for the colors).}
\label{fig:VN_Poisson}
\end{figure}

The matrix $M$ is then diagonalized to obtain the complex eigenvalues. These
eigenvalues are plotted in Fig. \ref{fig:VN_Poisson} for different Poisson 
ratio, with wave vectors of norm in $[0,\pi]$, directed by the unit vector 
$(1,1)^T1/\sqrt{2}$. The eigenvalues must to be located inside the stability 
disk (disk of complex number whose modulus is less than or equal to one)  for 
the numerical method to be stable for monochromatic disturbances.

\begin{figure}
\centering
\includegraphics[scale=0.4]{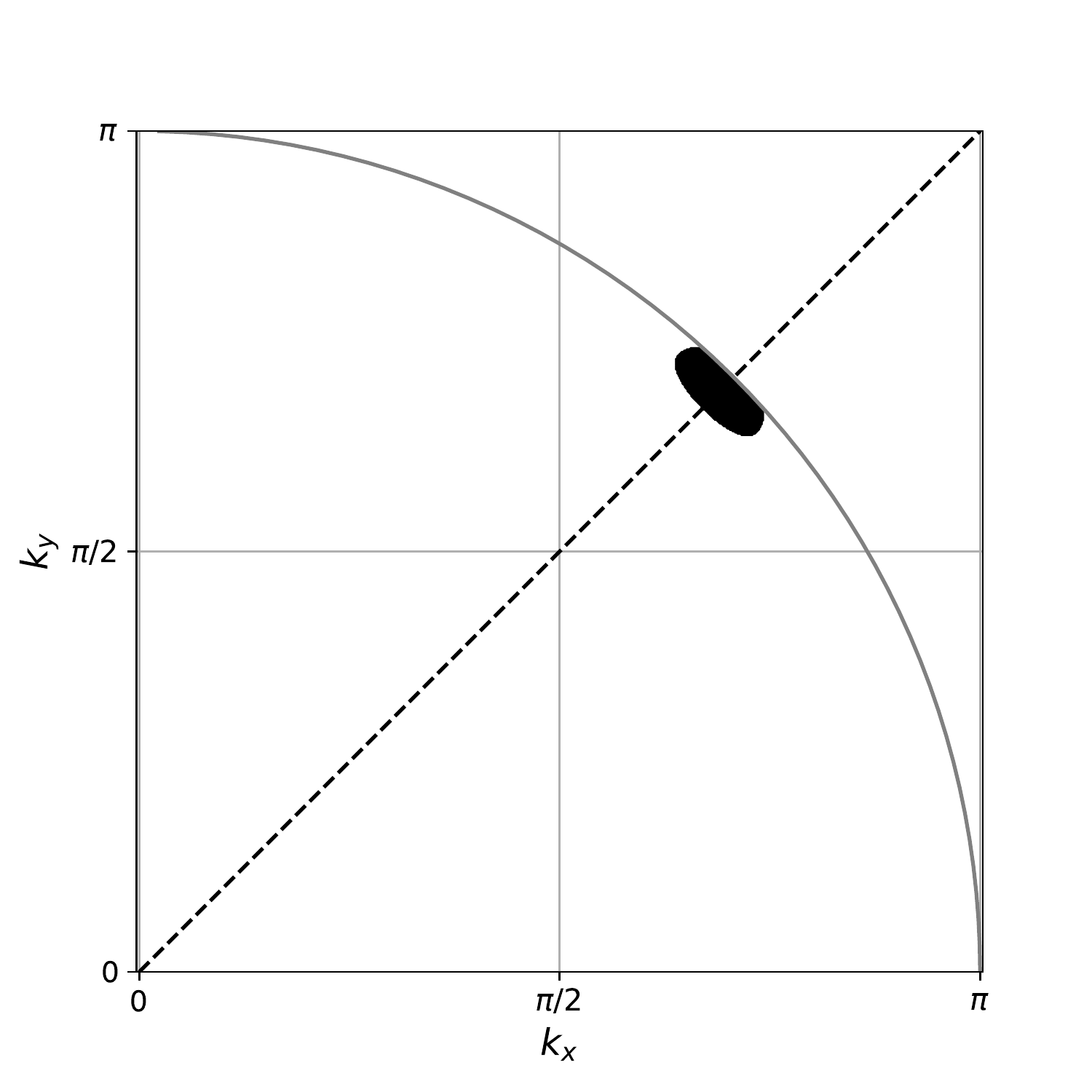}
\includegraphics[scale=0.4]{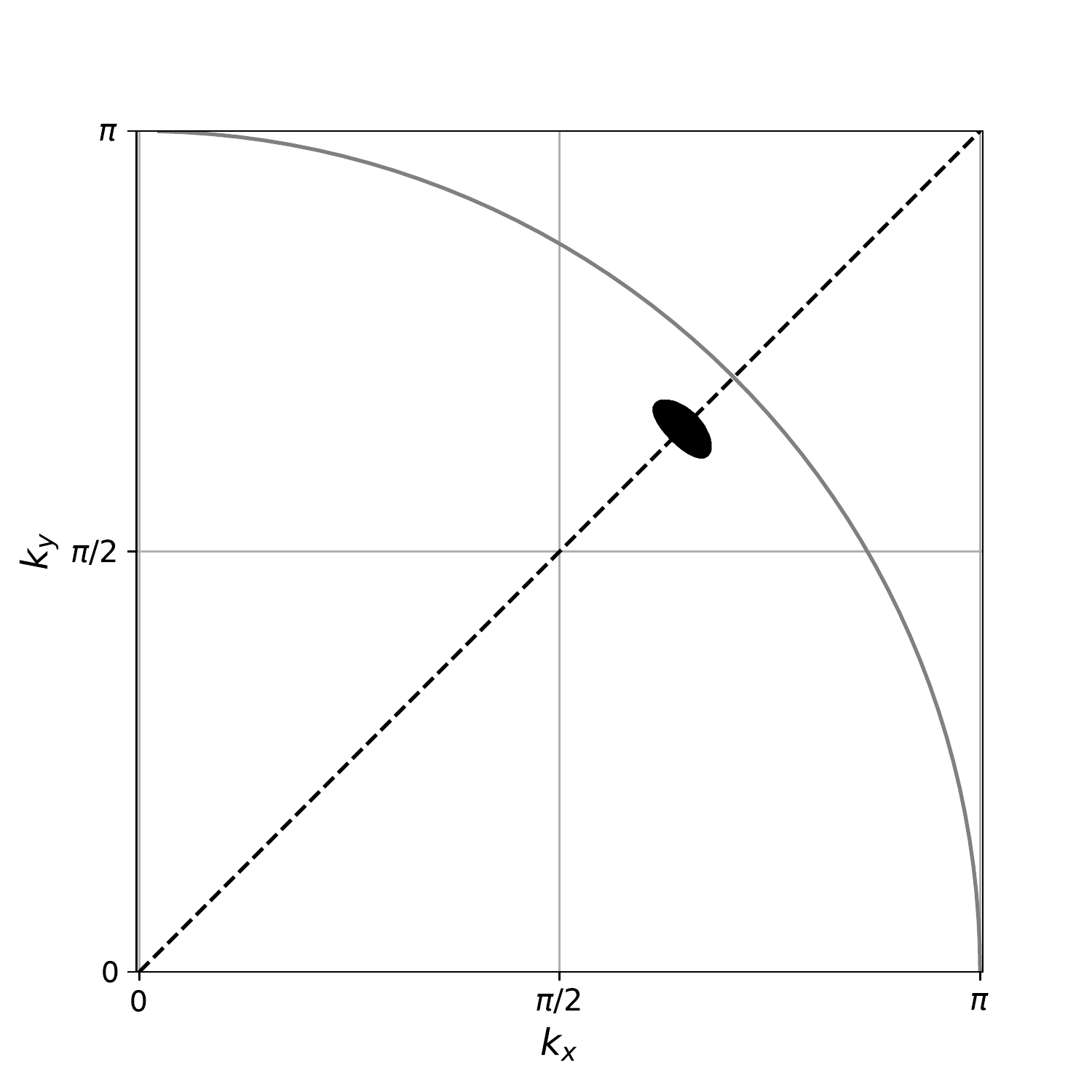}
\includegraphics[scale=0.4]{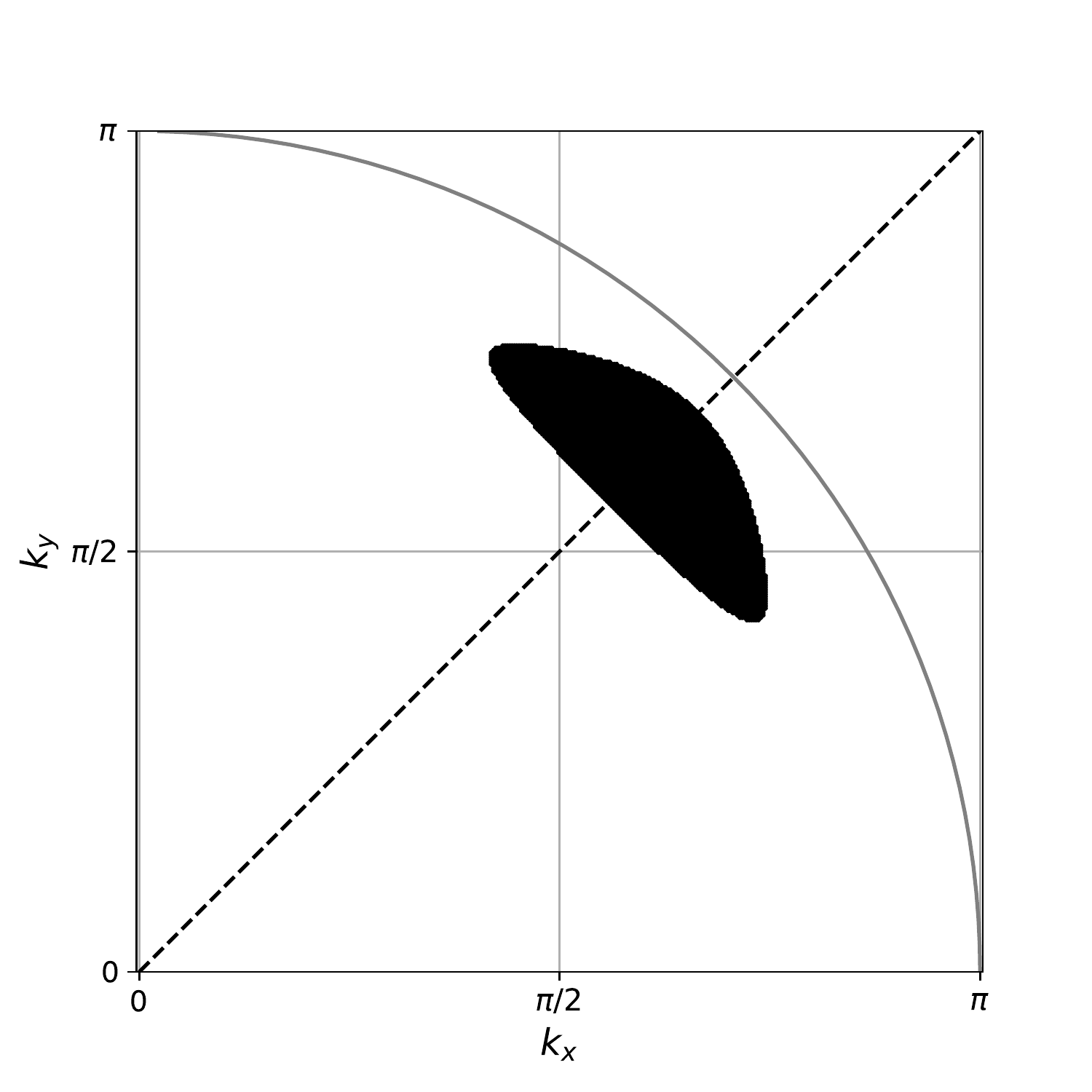}
\includegraphics[scale=0.4]{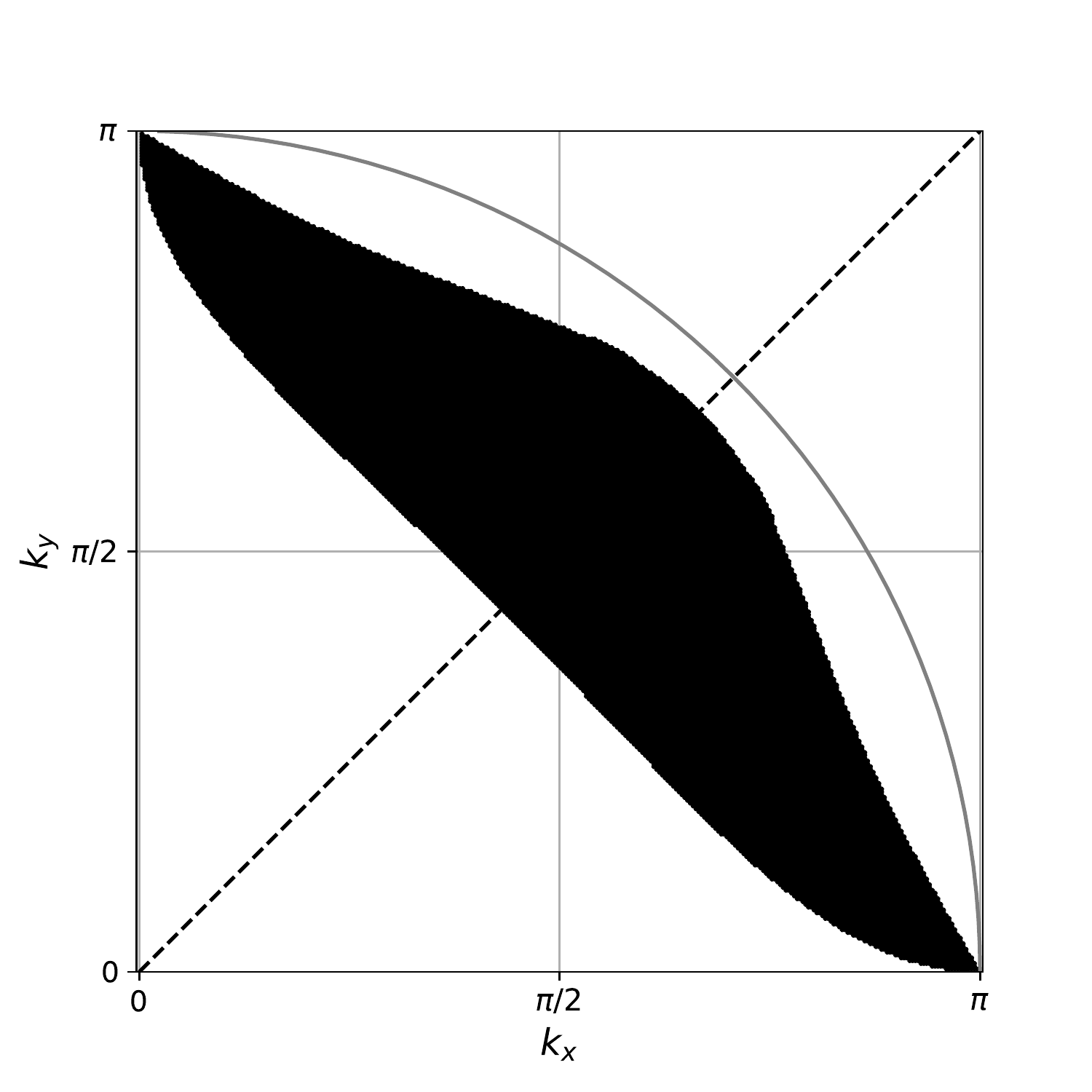}
\caption{Unstable wave vectors for different Poisson ratio.  Top-left: $\nu 
=0$; 
Top-
right: $\nu=0.3$; Bottom-left: $\nu=0.35$; Bottom-right: $\nu=0.4$. The case 
$\nu=0.25$ is not 
shown since all wave vectors are stable. Only one quadrant is displayed due 
to the elementary symmetries of the set.}
\label{fig:VN_Poissonk}
\end{figure}

We observe that for a 45° directed wave vector, for a Poisson ratio different
than $0.25$, there are some eigenvalues outside the stability disk, and 
particularly when the $\nu$ gets closer to the incompressible limit $0.5$. To 
investigate the other direction causing instability, we plotted on the 
$(k_x,k_y)$ plane, the points where some eigenvalues were outside the stability 
disk (i.e. with a modulus more than one). The results are 
shown in Figure \ref{fig:VN_Poissonk}.\\

\begin{figure}
\centering
\includegraphics[scale=1]{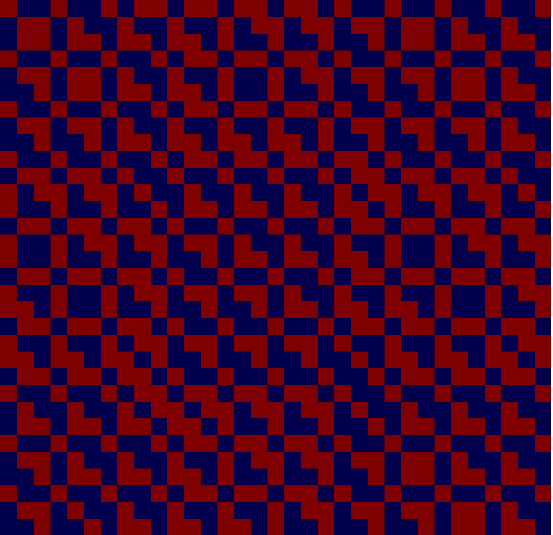}
\includegraphics[scale=1]{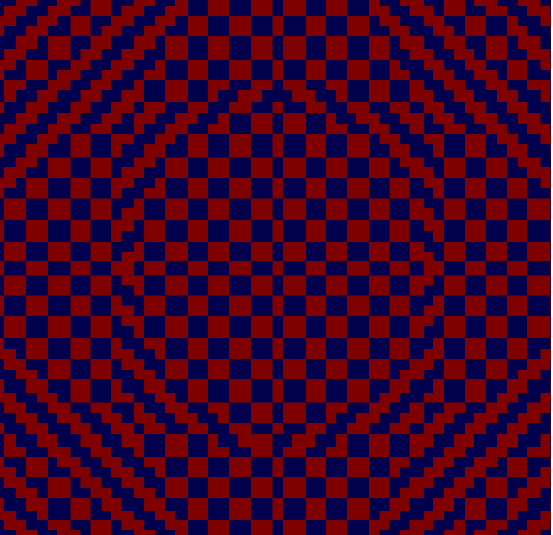}
\caption{Example of instability patterns. Left: 45°; Right: all-
directions. The color-map is saturated to blue-red colors due to great 
positive-negative values displayed: a sign of instability. In the 45° 
instability pattern, the diagonal stripes are spaces by $2.1$ spatial units. 
45°
patterns can be observed in low $\nu$ simulation after a long time, 
all-directions patterns can be observed in high $\nu$ simulation (here 
$\nu=0.4$).}
\label{fig:Instab_patterns}
\end{figure}

We see that for low Poisson ratio, the instability occurs only near the 45°
direction and for high wave vectors only (near the $\pi$-radius circle). 
This 
results in 45°-direction instability pattern (see figure 
\ref{fig:Instab_patterns}), after a long time (since only high wave-numbers are 
amplified, and since a Gaussian shape has very few high wave-numbers).

For high Poisson ratio, the domain of instability increases until all 
directions are unstable at high enough wave-numbers. This happens at $\nu=0.4$. 
Since we observe that instability occurs at lower wave-numbers for $\nu>0.25$ 
than for $\nu<0.25$, the high Poisson ratio simulations diverge more quickly.
But is is useful noticing that only very high wave vectors are unstable, 
typically $\vert \mathbf{k} \Delta x \vert > \pi/3$ even at $\nu =0.4$.
Therefore stable simulations can be recovered by refining the grid, or by 
filtering high frequencies at the end of a time step.

For a Poisson ratio $\nu>0.25$ the velocity of P waves is greater than the 
lattice speed:
\begin{equation}
    v_P(\nu>0.25)>\sqrt{3}v_S=1
    =\frac{\Delta x}{\Delta t}
\end{equation}
It is a classic source of instability to have an information supposed to 
propagate faster than the lattice speed. This is another way of understanding 
the instability of Poisson ratio greater than $0.25$.

In fact, due to the second-order centered finite difference in the calculation 
of $\partial_\alpha \rho$ in \eqref{FDrho}, the information can in reality 
propagate at speed 2 instead of 1. So, the absolute limitation of $\nu$ beyond 
which $v_P$ would faster than 2 is $\nu_{lim}=5/11\approx0.45$.

information is useful to understand that even with a stabilizing correction,  
the incompressible case $\nu=0.5$ is unreachable for regular sets with a  
definition of the "sound speed" with S waves velocity $b=v_S$.

Nevertheless, the present method can be used for practical simulations. It is 
unconditionnaly stable for Poisson solids, i.e. $\nu=0.25$, and can be used for 
finite-time simulations for $\nu \neq 0.25$ at classical LBM CFL number. 
Since the instability arises only at high wave numbers.

\section{Validation: Bulk wave propagation}
\label{sec:SPwaves}

As a first validation, one addresses  the propagation of both P and S waves in 
the bulk of a 2D linear isotropic elastic medium.
To this end, a $x$-force Gaussian source is placed in the middle of a domain 
periodic in $x$ and $y$ directions. The source is spatially Gaussian (instead of
point source for example) to avoid high spatial frequencies (high wave numbers) 
which behave less precisely and can lead to instability. The typical radius of 
the source is 4 $\Delta x$. This source is modulated through time with a classic
Ricker wavelet with a period of 20 $\Delta t$.

The D2Q9 simulation is compared with reference results obtained using a Fourier 
spectral method (with Crank-Nicolson time integration to allow  for CFL=1, i.e. 
$\Delta 
x=\Delta t$ to correspond with the LBM iterations) for several reasons:
\begin{itemize}
\item 2D analytical solutions of the wave equation are not 
trivial (they make use of Bessel's functions),
\item The source is not point source (which would involve convolutions for an 
analytical solution),
\item The simulation is periodic in both $x$ and $y$ directions.
\end{itemize}

The detail of this spectral method is explained in Appendix B.

\begin{figure}
\centering
\includegraphics[scale=1]{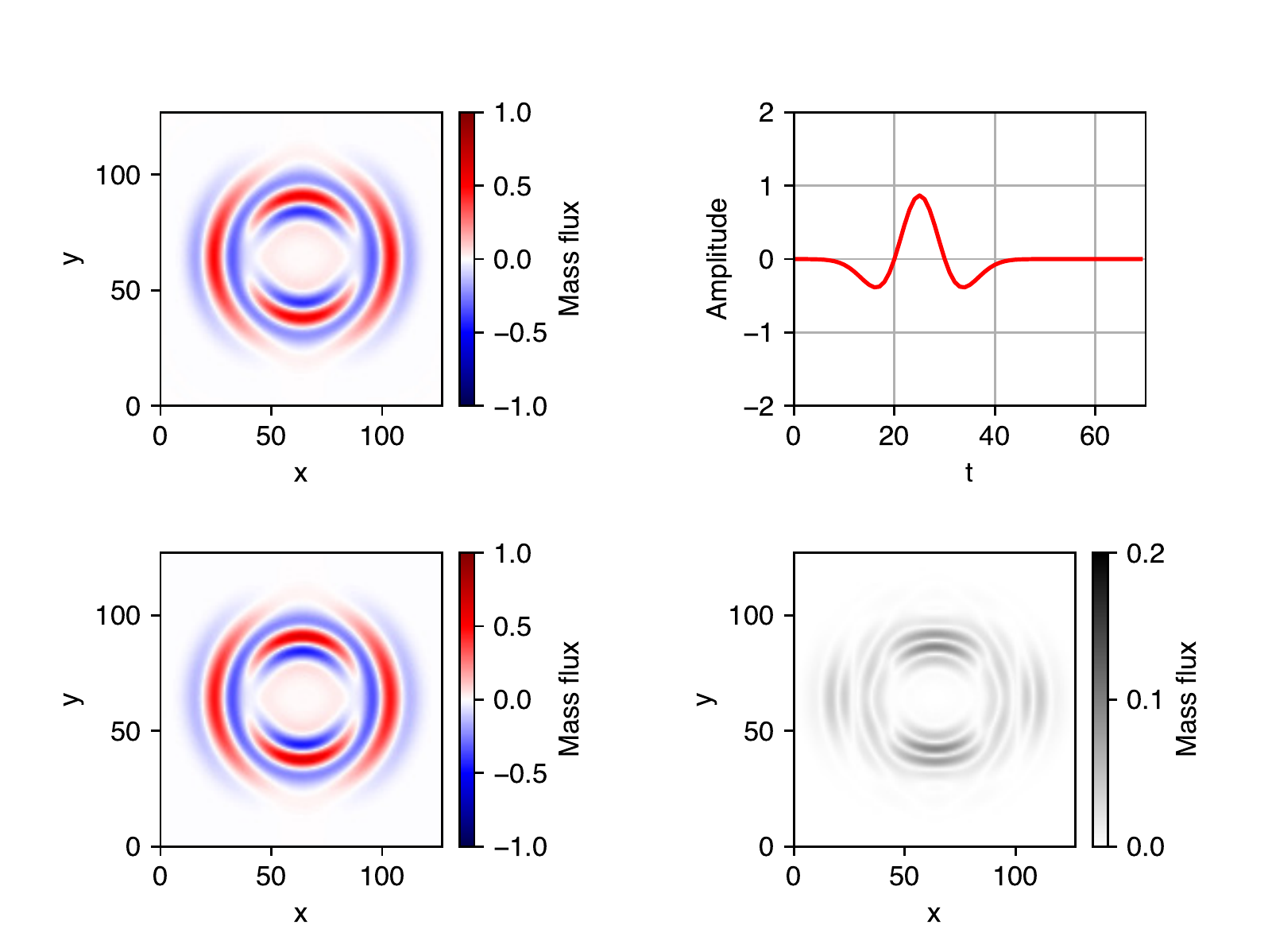}
\caption{Comparison of $j_x$ for LBM D2Q9 with the Fourier spectral method, for 
Poisson ratio $\nu=0.1$. Top-left: LBM D2Q9; Bottom-left: Fourier spectral; 
Top-right: Ricker wavelet modulation of the source; Bottom-right: absolute 
difference of $j_x$ between the two models.}
\label{fig:D2Q9_Spectral}
\end{figure}

The simulations are performed on 64x64, 128x128, 256x256 and 512x512 grids 
using 
$\tau=0.55\Delta t$  along
with $\nu= 0.0,\ 0.1,\ 0.2,\ 0.25$ and $0.3$. Instantaneous results 
obtained at time $70 \Delta t$ on the 128x128 grid for $\nu=0.1$ are 
displayed in Fig. \ref{fig:D2Q9_Spectral}.
One can see  that both LBM and spectral method capture the behaviour of P waves
and S waves very accurately. The P waves are the ones on the left-right sides 
(the fastest 
travelling waves) and the S waves are the one on the top-bottom sides.

\begin{figure}
\centering
 \includegraphics[scale=0.5]{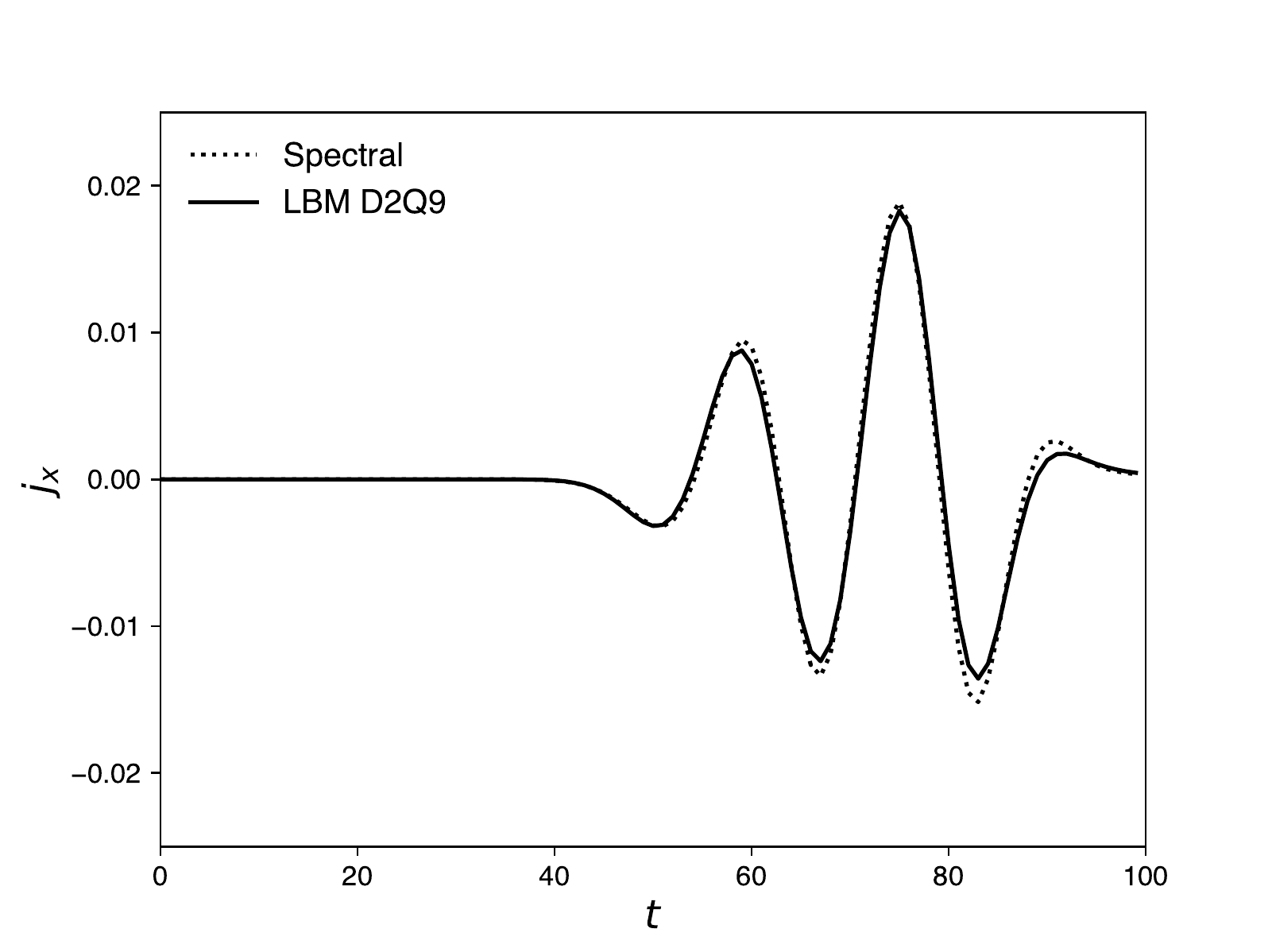}
\caption{Comparison of the seismogram of the longitudinal mass flux  $j_x$ for 
LBM D2Q9 with the spectral 
method, for a Poisson ratio $\nu=0$.}
\label{fig:Seismogram}
\end{figure}

In order to quantify the difference between the LBM and the spectral method, a 
seismogram of a particular station (here $x=2 N_x /3$,$y=2 N_y /3$ ) is plotted 
in Fig. 
\ref{fig:Seismogram}. It is observed that a very satisfactory agreement is 
recovered, even on a coarse grid for the LBM computation.

The accuracy of the LBM method is further investigated by measuring its 
effective order of convergence.
The relative $L_2$ norm of the error committed on the longitudinal mass flux 
$j_x$ is defined as

\begin{equation}
    p=\sqrt{\frac{\sum_{i,j=1,1}^{N_x,N_y} \left(j_x^{\rm{LBM}} 
(i,j)-j_x^{\rm{spectral}}(i,j)\right)^2}
    {\sum_{i,j=1,1}^{N_x,N_y}  \left(j_x^{\rm{spectral}}(i,j)\right)^2}}
    \label{pmisfit}
\end{equation}

\begin{figure}
\centering
\includegraphics[width = 0.7 \textwidth]{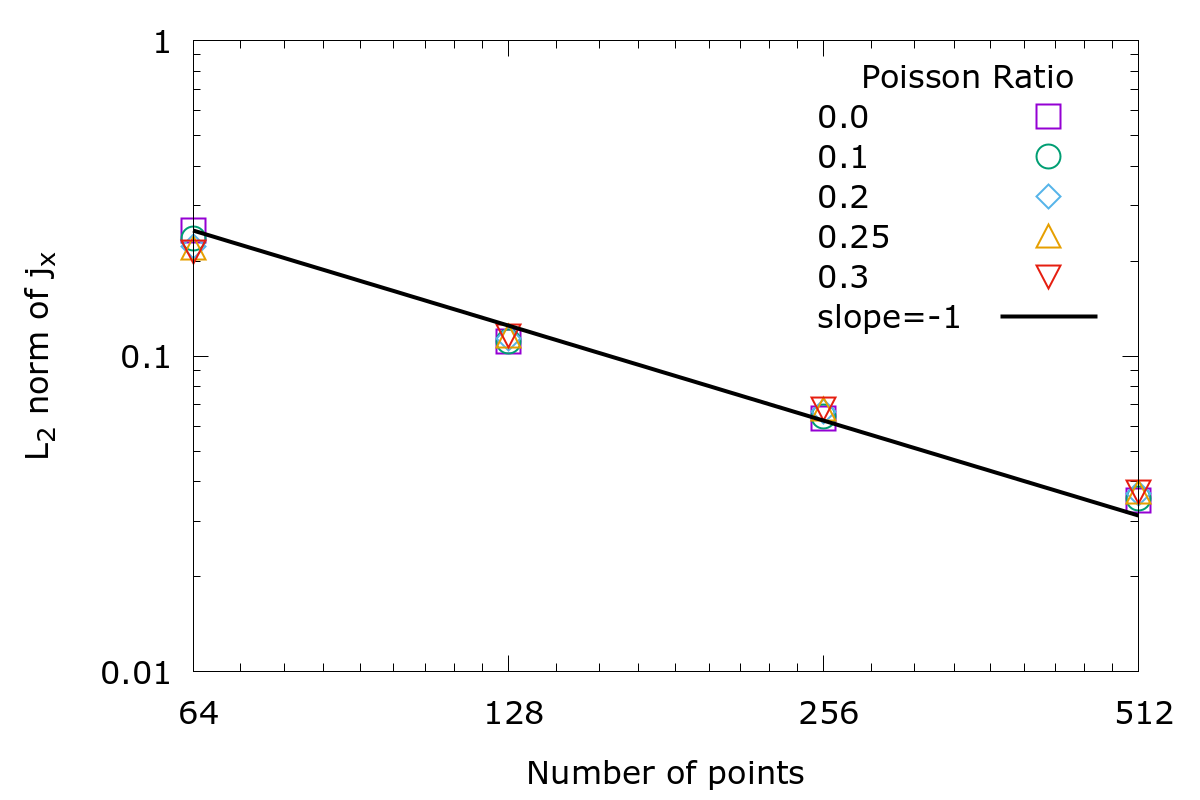}
\caption{Grid convergence of the $L_2$ norm of the error on the longitudinal 
mass 
flux $j_x$.}
\label{Fig:order-of-convergence}
\end{figure}

Values computed are given in Table \ref{table:errorL2} and plotted in Fig. 
\ref{Fig:order-of-convergence}.
It is observed that the error exhibits a weak sensitivity to the Poisson ratio, 
which is a good property, and that a first order of convergence is obtained on 
the mass flux.

\begin{table}
\caption{$L_2$ norm of the error computed on the longitudinal mass flux $j_x$ 
for several combinations of the grid resolution $Nx \times N_y$ and the Poisson 
ratio $\nu$.}
\begin{center}
\begin{tabular}{|c|c|c|c|c|c|}
\hline
 & $\nu =0$ & $\nu =0.1$ & $\nu =0.2$ & $\nu =0.25$ & $\nu =0.3$ \\ \hline
 $N_x = N_y =$ 64 & 0.2508 & 0.2351 & 0.2227 & 0.2185 & 0.2141 \\ \hline
 $N_x = N_y =$ 128 & 0.1112 & 0.1111 & 0.1137 & 0.1155 & 0.1164 \\ \hline
 $N_x = N_y =$ 256 & 0.0633 & 0.0644 & 0.0667 & 0.0678 & 0.0681 \\ \hline
$N_x = N_y =$  512 & 0.0349 & 0.0355 & 0.0366 & 0.0371 & 0.0371 \\ \hline
\end{tabular}
\end{center}
\label{table:errorL2}
\end{table}%

\begin{figure}
\centering
\includegraphics[scale=0.5]{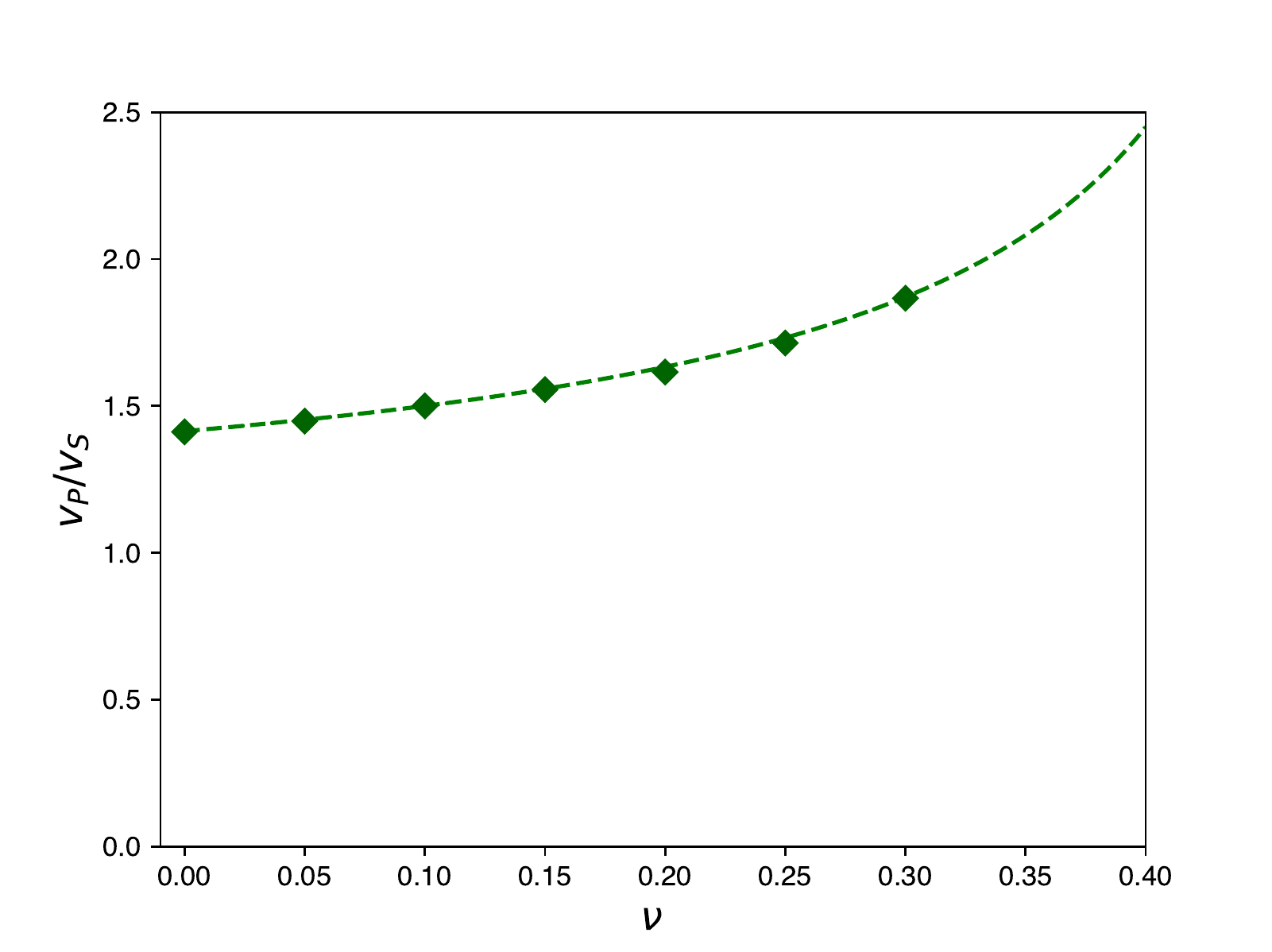}
\caption{Comparison of the theoretical (dashed line) and LBM D2Q9 results 
(diamond markers) relation between  $\nu$ and $v_P/v_S$}
\label{fig:vPvS}
\end{figure}

To check the velocity of the bulk waves, some measurement of the travel time of 
P waves (time to reach the left-right borders from the center) and S waves 
(time to reach the top-bottom borders from the center) have been performed  for 
several values of the Poisson ratio(see figure \ref{fig:vPvS}). The maximum 
relative error is 1.1\% on the 128x128 grid, corresponding to a very accurate 
capture of the wave 
physics.

The present results show that the D2Q9 Lattice Boltzmann method yields very 
accurate prediction of bulk P and S waves, with an accuracy similar to those 
reported with other methods, e.g. \cite{obrien2009, obrien2011, 
obrien_lattice_2012, xia_modelling_2017}.

\section{Boundary conditions}
\label{sec:BCs}
The developments given in \cite{jmurthy_lattice_2018} and in the above section 
were focused on the bulk wave propagation in infinite media, and only periodic 
boundary conditions were used. We now extend the method by deriving boundary 
conditions for more realistic cases: rigid interfaces with null displacement, 
free surface and non-reflecting outlet boundary condition.

\subsection{Rigid wall boundary conditions}
The simplest BC is probably the "Bounce-Back" boundary condition (BB). 
For fluids it represents the \emph{no-slip condition}, typically used to
implement a steady solid wall in a viscous fluid, leading to (for instance for 
a bottom wall): $v_{x,wall}=0$ with $v$ the velocity. For solids, a rigid 
interface interface would mean $j_{x,wall}=0$, i.e.  the displacement and the 
mass flux are null on the boundary. This would exist with the interface of our 
solid of interest with a rigid wall (for instance another solid with a much 
greater density or stiffness). The usual BB condition imposes the wall location 
to lie between two nodes : 
\begin{equation}
    y_{wall}=y_b+\frac{\Delta y}{2}\vec{n}
\end{equation}
with $y_b$ the last layer near the interface and $\vec{n}$ the normal interface 
unit vector pointing outwards the solids. Therefore, the present Bounce Back 
condition is implemented as 
\begin{equation}
    f_{\bar{i}}(x_b,t+\Delta t)=f_i(x_b,t)
    \label{BB}
\end{equation}
with $c_{\bar{i}}$ being the opposite of $c_i$: $c_{\bar{i}}=-c_i$

\subsection{Free surface boundary condition}
We consider now an interface with a medium with a much lower density and
stiffness than the solid, e.g. gas or void. In the following, the case of the 
interface between the solid and air is used for the sake of clarity.

The solid/air interface location is $\Delta x/2$ far the last LBM node $x_b$
located in the solid medium:

\begin{equation}
x_{air}=x_b+\frac{\Delta x}{2}\vec{n}
\end{equation}
with $\vec{n}$ the normal interface unit vector pointing outwards the solids.
The bouncing condition is now :

\begin{equation}
f_{\bar{i}}(x_b,t+\Delta x)=-f_i(x_b,t)+2w_i\left(\rho_{air}+\frac{1}{2b^4}
P^n_{\alpha\beta,air}\left(c_{i\alpha} c_{i\beta}-b^2\delta_{\alpha\beta} 
\right) \right)
\end{equation}

where $\rho_{air}$ is the density estimated at the border with a linear
interpolation

\begin{equation}
\rho_{air}=\frac{3}{2}\rho_{b}-\frac{1}{2}\rho_{in}
\end{equation}
where $\rho_{in}$ is the density one node further inside the solid
\begin{equation}
x_{in}=x_b-\Delta x\vec{n}
\end{equation}

The $P_{\alpha\beta,air}^n$ tensor at the interface is :
\begin{equation}
P_{\alpha\beta,air}^n=P_{\alpha\beta,air}-b^2\rho_{air}\delta_{\alpha\beta}
\end{equation}

To explain how to find $P_{\alpha\beta,air}$, we consider  the case of a 
vertical interface:
\begin{equation}
P_{\alpha\beta,air} =
\begin{pmatrix} 
P_{xx,air} & 0\\
0 & 0
\end{pmatrix}
\end{equation}

because $P_{\alpha\beta,air}n_\beta=F_\alpha=0$,  $F_\alpha$ is the force 
applied by  air on the solid at the interface, which is assumed to be negligible
in usual pressure and temperature conditions. $P_{xx,air}$ is obtained with 
linear interpolation similar to $\rho_{air}$.

\begin{figure}
    \centering
    \includegraphics[scale=0.5]{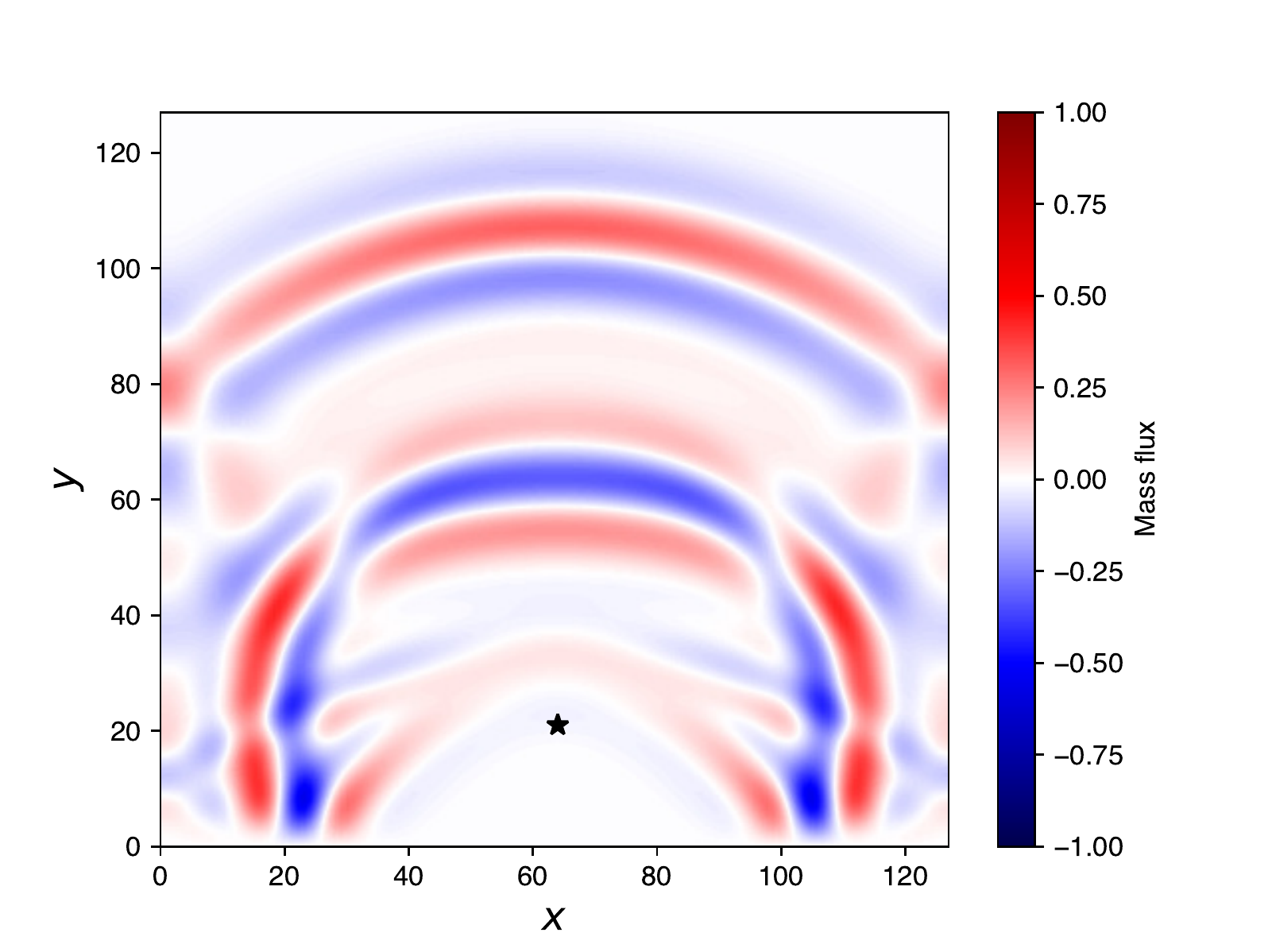}
    \includegraphics[scale=0.5]{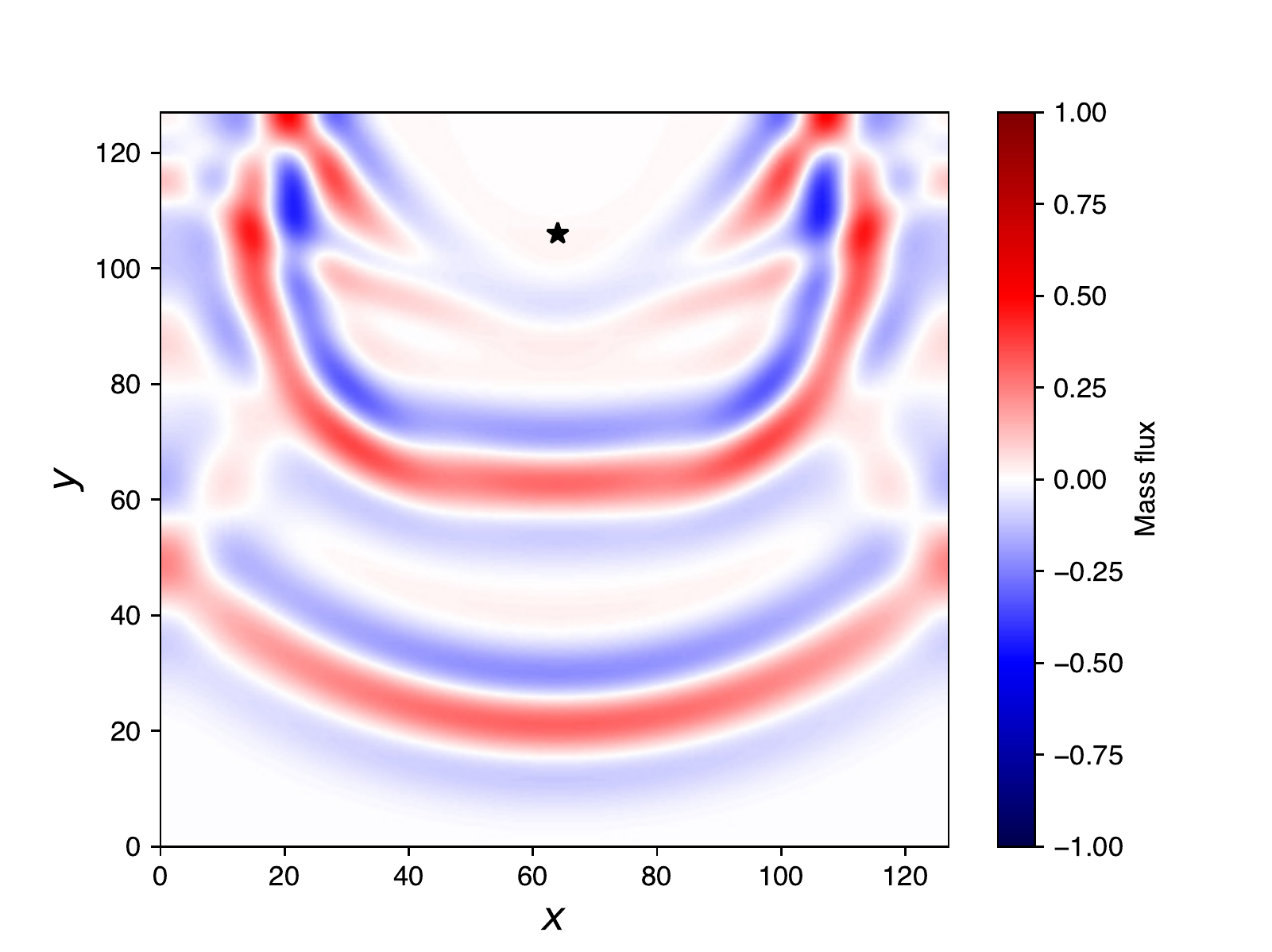}
    \caption{Test of the BCs: reflection of a vertical P wave on different 
    interfaces. Left: bottom rigid wall (BB); Right: top free surface (ABB).
    The star marker indicates the position of the source}
    \label{fig:BB_Free}
\end{figure}

\subsection{Validation of Rigid Wall and Free Surface boundary conditions}

We now validate the two boundary conditions discussed above by considering wave 
reflection on these two kinds of boundaries. 

\begin{figure}
\centering
\includegraphics[scale=0.3]{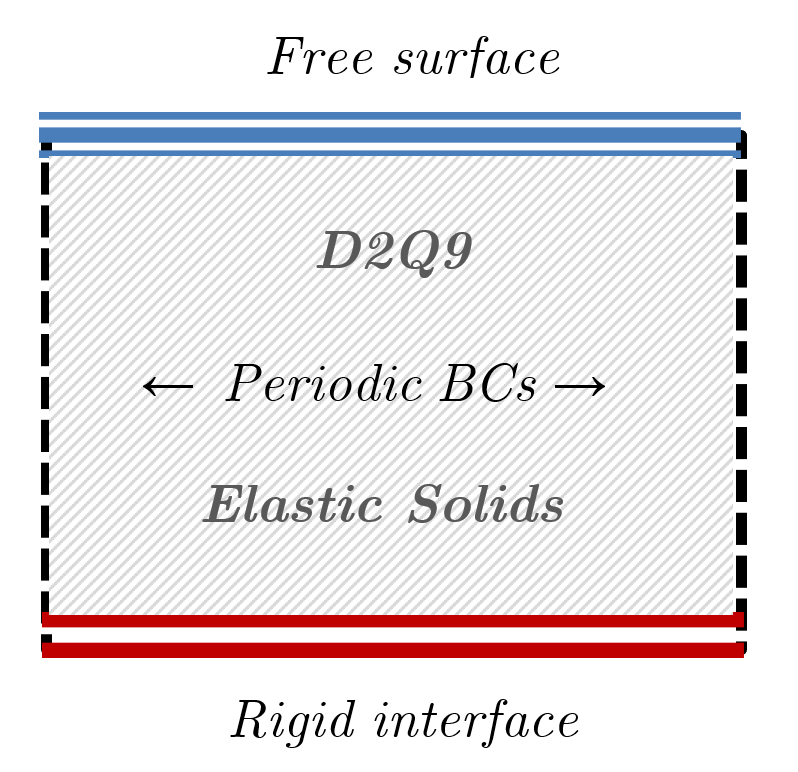}
\caption{Configuration of the first BC test}
\label{fig:ConfigBC}
\end{figure}

To this end, a domain with periodic lateral conditions is defined, the BCs of 
interest being implemented on the top (free surface case) or bottom boundary 
(rigid wall case), as illustrated in Fig. \ref{fig:ConfigBC}.
A directional source pointing vertically is added near the interface of 
interest. We observe that the waves are bouncing off these interfaces. The 
larger circle arc is the first P-wave with a direct path and the second 
concentric circle is the P-wave reflected.

Two criteria indicate that the boundary conditions are accurately implemented. 

The first one deals with the sign of the velocity.
For a free surface, there is no change in the sign of the speed (i.e. phase 
change) after the reflection, since the condition is a Neumann condition for 
velocity. For the rigid interface, there is a change in the sign of the speed 
(i.e. phase inversion) because it is a Dirichlet condition on velocity.
This can be understood considering the following simplified mass flux equation:

\begin{equation}
\partial _t j_\alpha + \partial _\beta P_{\alpha \beta} = 0
\end{equation}

Considering   an horizontal boundary at $y=0$, it leads to $\partial_t j = 
-\partial_y P$, yielding $P(c_s t \mp y)=\pm c_s j(c_s t \mp y)$, so that $P=0$ 
at the border is equivalent to $P(0,t)=P_i(c_s t - 0) + P_r(c_s t + 0) = 0$, 
where the i and r subscripts are related to incident and reflected waves, 
respectively.
For rigid walls with $j=0$, the Dirichlet condition gives the result in a 
straightforward way.

The second criterion is the speed magnitude at the border. For a free surface 
the incident and reflected wave interfere constructively while for the rigid 
interface, the interference is destructive. This is visible with a darker 
(respectively lighter) area near the border.

\subsection{Absorbing layers for non-reflecting boundary conditions}
To simulate a semi-infinite medium, we need to introduce non-reflecting  
boundary conditions in order to define outlet boundary conditions.
Different approaches have been tested for fluids, based either on the use of 
improved BCs on the outlet plane or the implementation of sponge layers in which
fluctuations are damped, or a combination of these two approaches. 
Numerical experiments performed in fluid LBM framework have shown that the use 
of a sponge layer is mandatory to get very accurate results for internal flows 
or on small computational domain.

Contrary to fluids where the viscosity can be changed with $\tau$ to introduce 
a progressive damping effect, the Chapman-Enskog-like analysis given above 
shows that $\tau$ doesn't induce a viscous behaviour in the present case. 
Therefore, in the present solid LBM case, it is chosen to design absorbing 
layers. The principle is to smoothly increase a damping factor near the borders 
to completely remove the reflected waves. This is done by adding a linear 
penalty term in the mass flux equation:

\begin{align}
    \begin{split}
 &\partial_{t}\rho + \partial_{\alpha}j_{\alpha} = 0\\
 &\partial_{t}j_{\alpha} + \partial_{\beta}P_{\alpha\beta} =
 \frac{\mu-\lambda}{\rho_0} \partial_\alpha \rho
 -A j_\alpha\\
 &\partial_{t}P_{\alpha\beta} + \partial_\gamma 
 Q_{\alpha\beta\gamma}^{eq} =0
    \end{split}
    \label{MomentChainAbsorbing}
\end{align}\\
where $A$ is the damping factor and the term $-Aj_\alpha$ is treated as a
forcing (exactly like an external forcing or  the elastic artificial forcing).

The associated modified Navier equation is :

\begin{equation}
\partial_t^2 j_\alpha + A\partial_tj_\alpha= \frac{\lambda+\mu}{\rho_0}
\partial_\alpha \partial_\beta j_\beta + \frac{\mu}{\rho_0} \partial_\beta^2 
j_\alpha + \frac{1}{\rho_0}\partial_t F_\alpha
\label{RelaxationNavierJ}
\end{equation}

A plane-wave analysis shows that all modes are damped in the absorbing  layer.

\section{Validation: Rayleigh surface waves}
The last validation case is related to the simulation of surface waves, that are
very important in seismology. The motion being restricted in the plane in the 
present paper (2D simulations), the only surface waves possible are the Rayleigh
waves (R waves). They stem from the combination of both P and S waves and their 
reflections at the surface. Their relative speed with P and S waves are also 
varying with the Poisson ratio but they are slightly slower than S waves. In the
particular case of Poisson solids ($\nu=0.25$), $v_R=v_S\sqrt{0.8453}\approx 
0.92v_S$ \cite{aki_quantitative_2009}. Rayleigh waves have an amplitude that 
decays exponentially with depth. The motion of the particle is ellipsoidal, like
the the motion of particle of fluids in the swell. Rayleigh waves have been 
reported to be accurately captured using ELM (see 
\cite{vallegarcia2003,obrien2014} and references given herein) but, to the 
knowledge of the authors, they have not been computed via LBM up to now.

To isolate and observe surface waves, the following numerical experiment has be 
conducted on a 100x300 computational grid. A free surface condition is
implemented on the top boundary and the three remaining borders are treated with
absorbing layers with a thickness of 30 nodes each. 
\begin{figure}
\centering
    \includegraphics[scale=1]{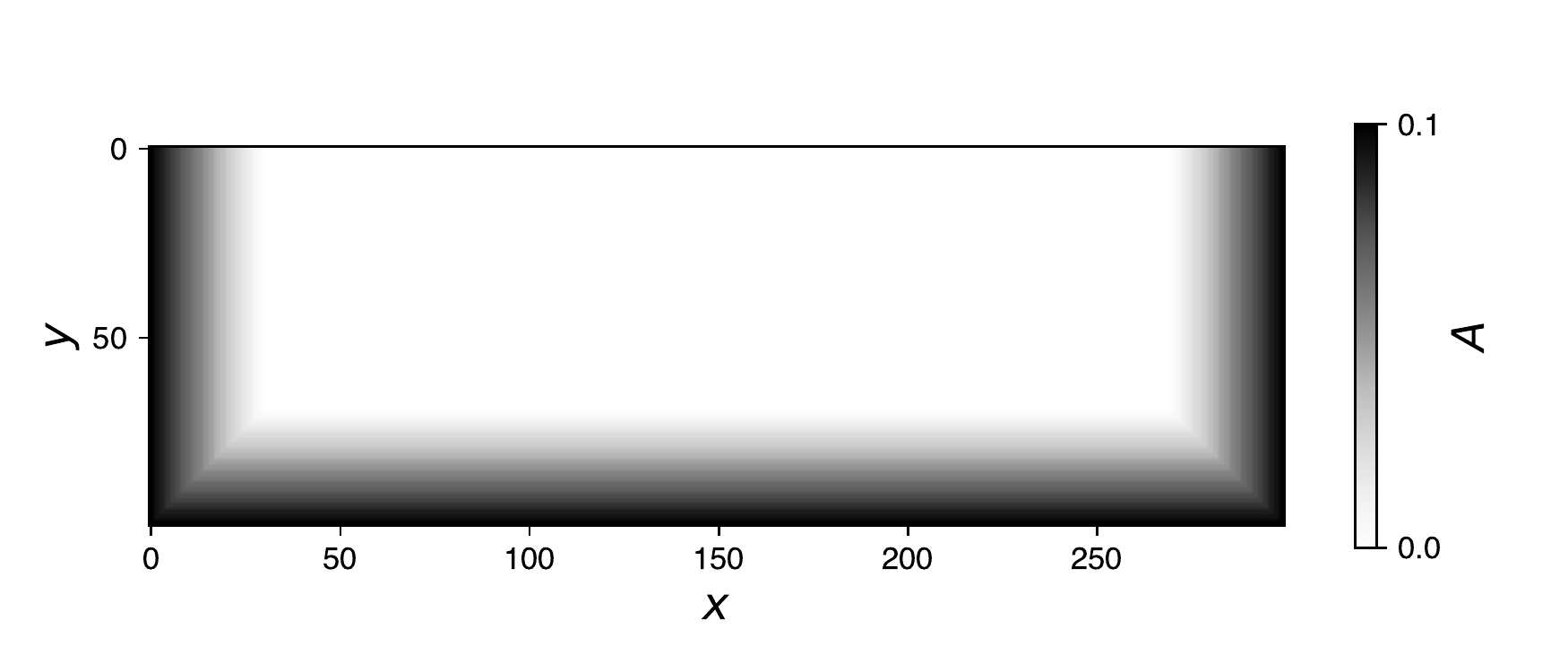}
    \caption{Absorbing layers}
    \label{fig:3AbsorbingLayers}
\end{figure}

As a first step, absorbing layers are used to damp bulk S and P waves in order 
to isolate Rayleigh surface waves. In the second step, absorbing layers are 
removed and periodic conditions are imposed on lateral boundaries in order to 
let the Rayleigh waves propagate freely. The source is located on the upper left
corner so that all waves going to the left are quickly removed (absorbed by the 
left sponge layer). The fastest wave (P waves) going to the right are also 
removed by the right absorbing layer, then (after 400 time steps) the left-right
absorbing layers are replaced by a simple left-right periodic condition. 

A 2D circular bulk wave with constant energy has a typical amplitude decaying as 
$1/\sqrt{r}$ while a surface wave 
in the present case will be a 1D wave, whose amplitude will be constant in the 
energy-preserving case. 
Thanks to that difference, surface waves should last longer than the  S 
waves that are travelling to the right just slightly faster). 

\begin{figure}
\centering
\includegraphics[trim=0cm 0.5cm 0cm 1.8cm,clip,scale=0.6]{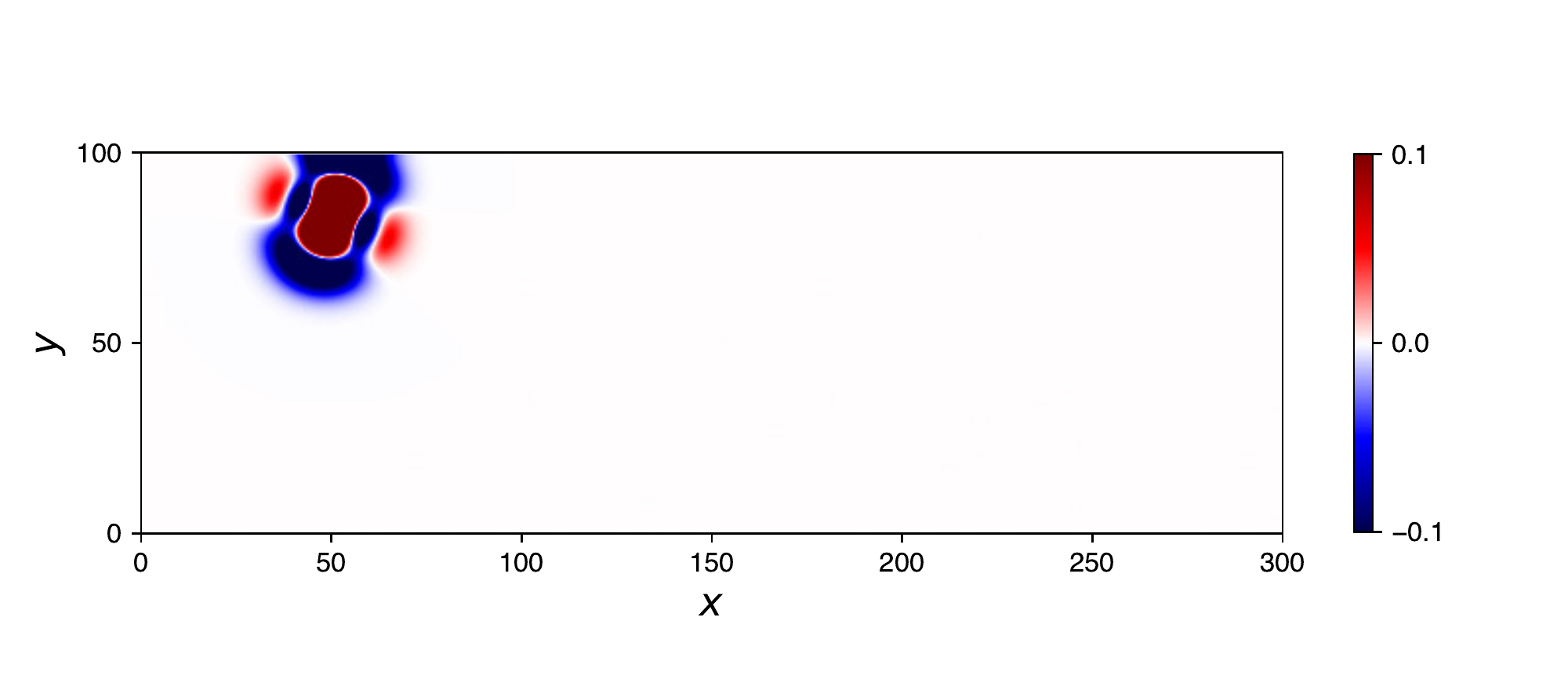}
\includegraphics[trim=0cm 0.5cm 0cm 1.8cm,clip,scale=0.6]{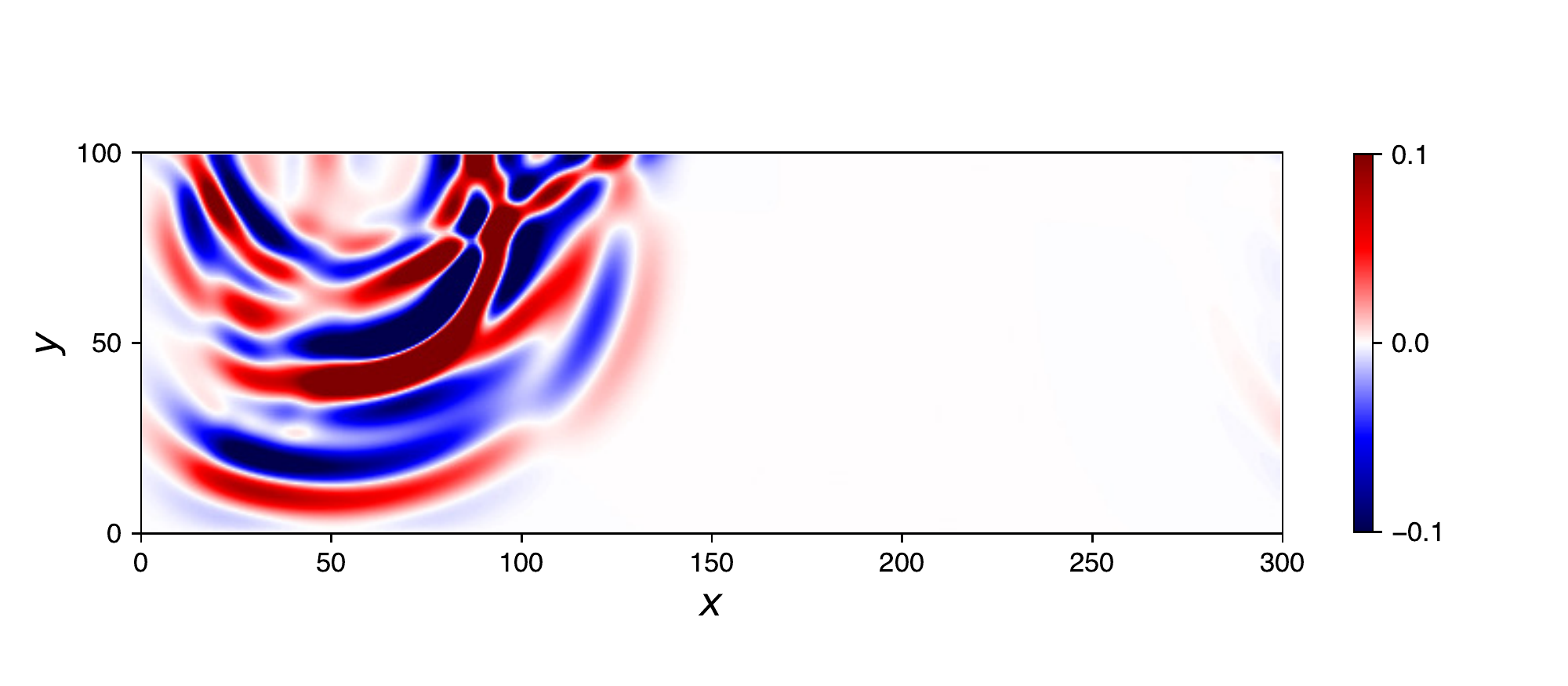}
\includegraphics[trim=0cm 0.5cm 0cm 1.8cm,clip,scale=0.6]{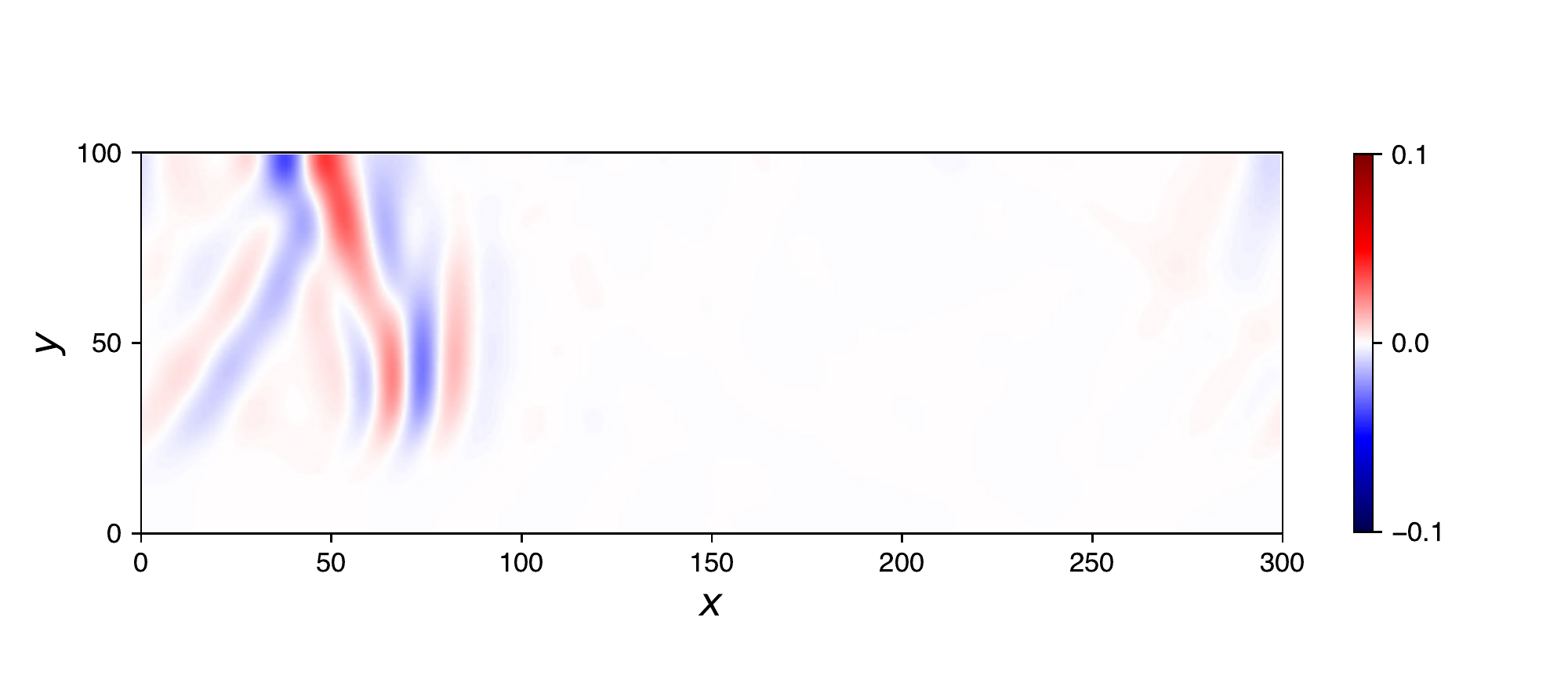}
\includegraphics[trim=0cm 0.5cm 0cm 1.8cm,clip,scale=0.6]{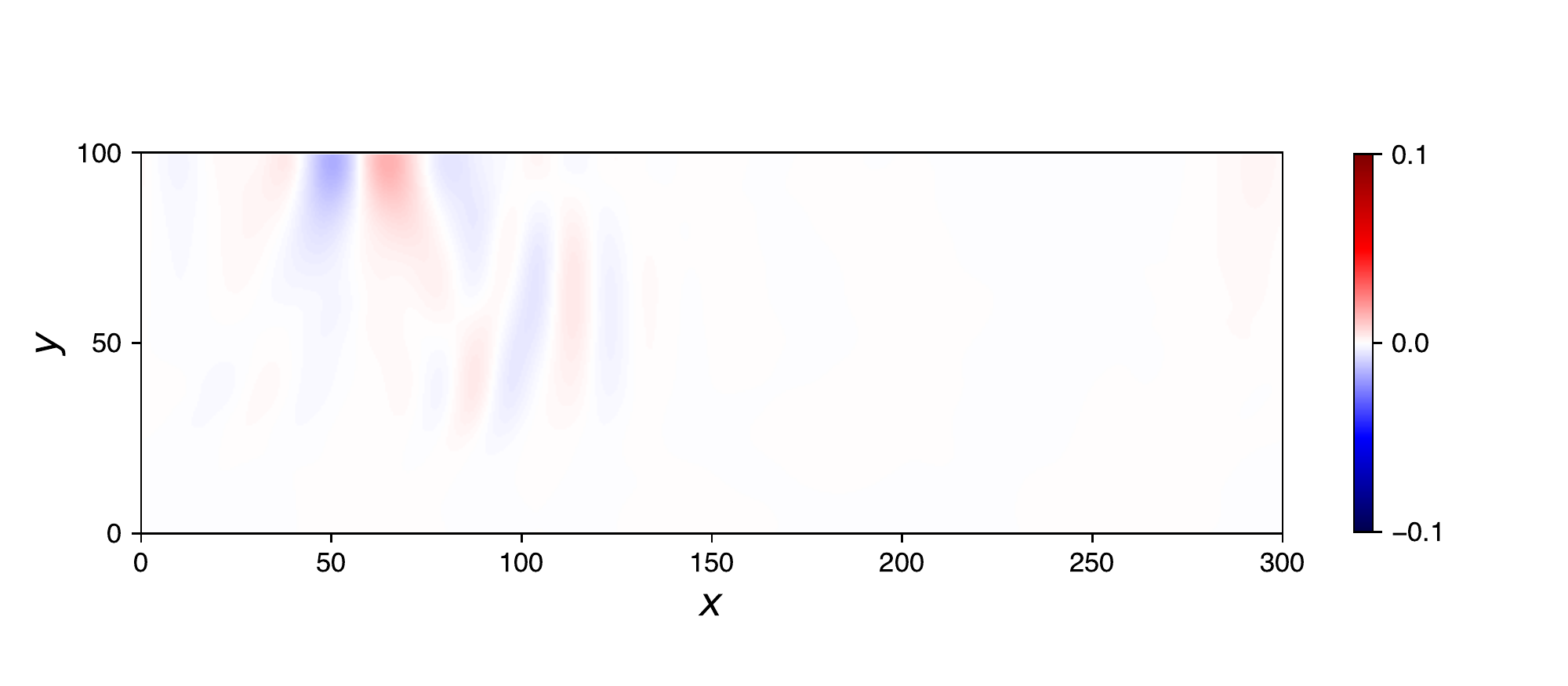}
\includegraphics[trim=0cm 0.5cm 0cm 1.8cm,clip,scale=0.6]{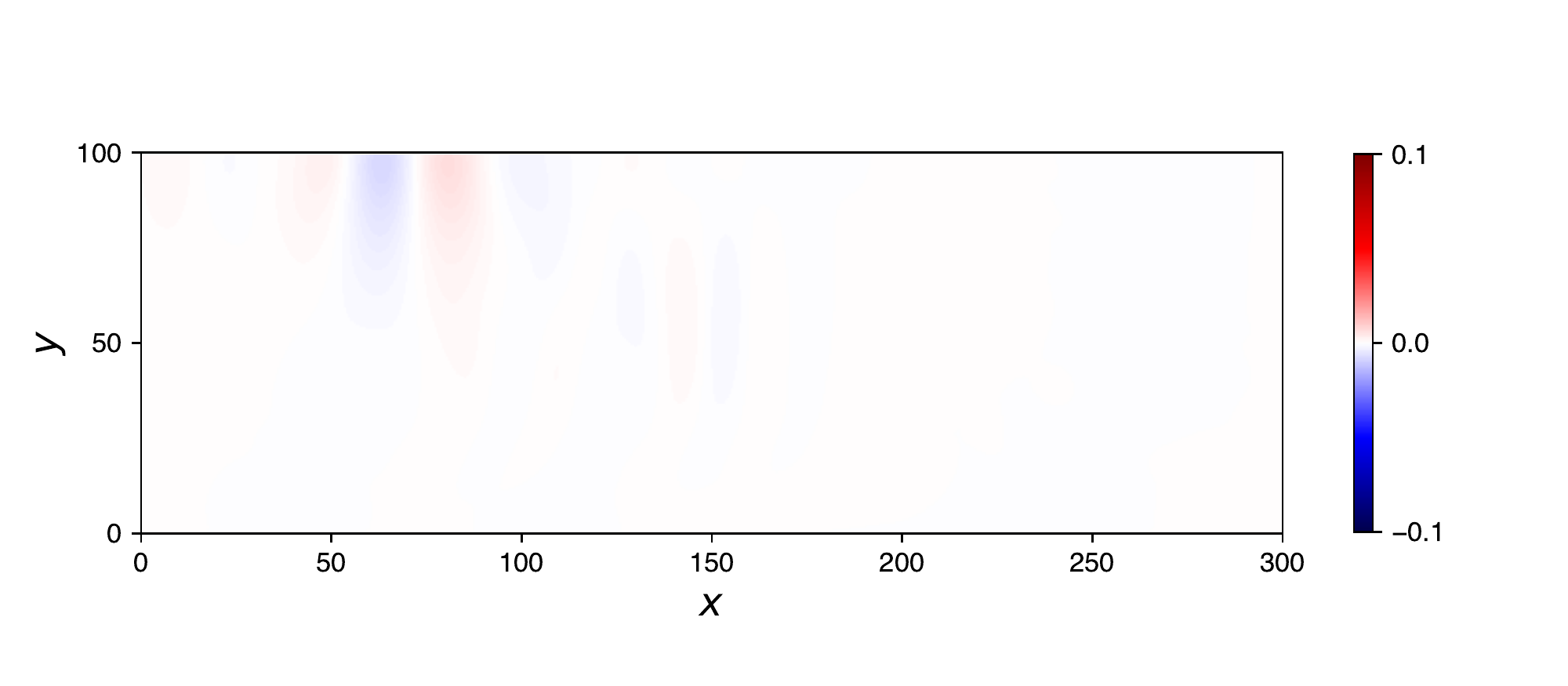}
 \caption{Vertical mass flux profile at different time, for $\nu=0.25$. The 
 star marker indicates the position of the source. From top to bottom: iteration
30, 100, 600, 1200, 1800}
 \label{fig:Rayleigh}
\end{figure}

The results are displayed in Fig. \ref{fig:Rayleigh}. The simulation 
qualitatively captures the behaviour of Rayleigh waves. At time steps 600 and 
1200 there remains some S waves also going to the right because they are only 
slightly faster than Rayleigh waves (at $\nu=0.25$,  $v_R\approx92\%v_S$). The 
speed of R waves have been measured with a 0.14\% relative error, by measuring 
the number of time steps needed to achieve for two complete turns around the 
periodic surface: 
\begin{align}
    \begin{split}
        v_R^{\rm{measured}}=0.5315, \quad 
v_R^{\rm{theo}}=\sqrt{\frac{0.8453}{3}}=0.5308
    \end{split}
    \label{vR}
\end{align}

However, the amplitude of these Rayleigh waves is slowly decreasing. This is
consequence of the error term found in Eq.  \eqref{MomentChain2} and underlined 
in the Von Neumann analysis.

\begin{figure}
\centering
\includegraphics[scale=0.6]{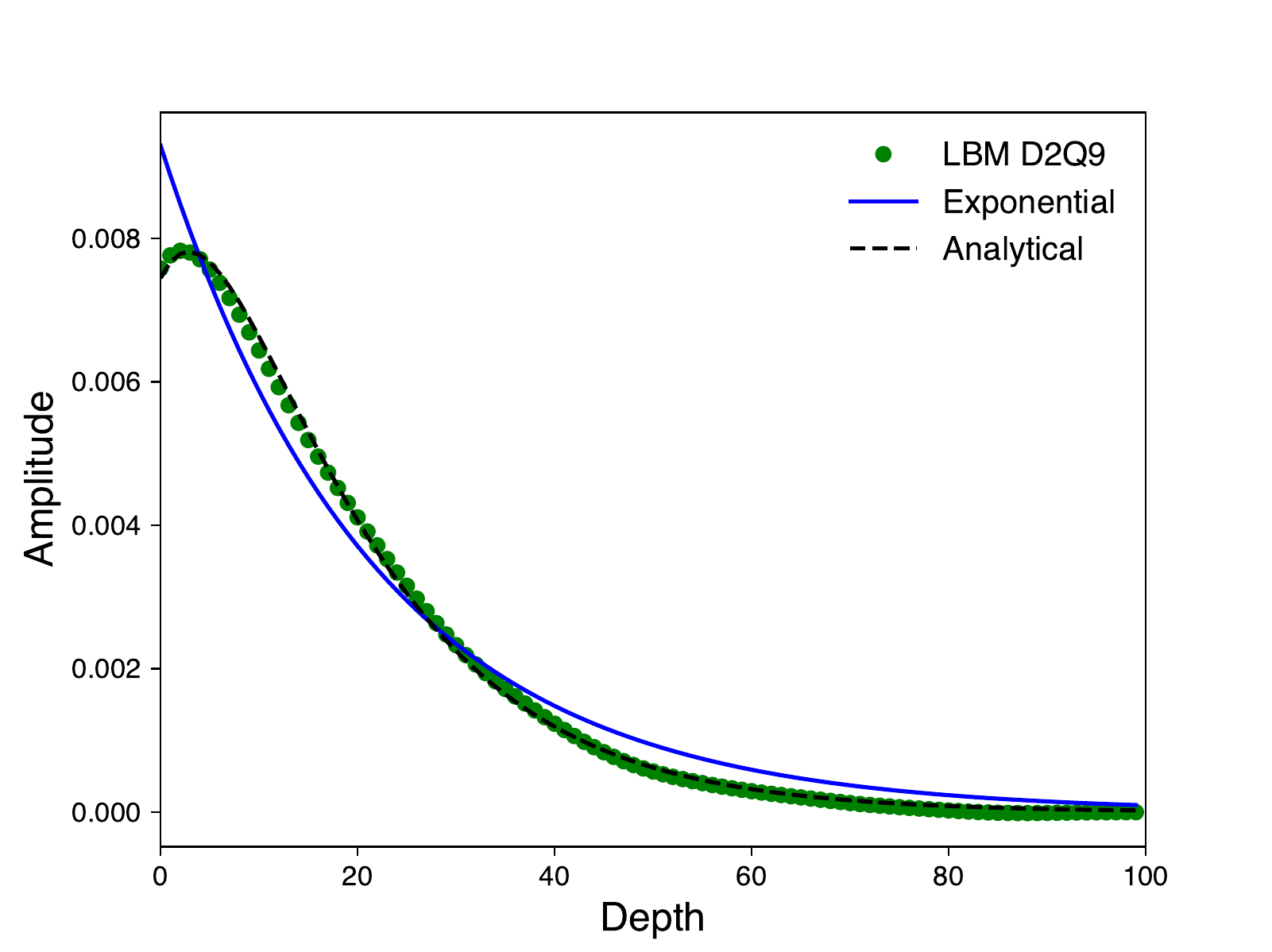}
\caption{Profile of the amplitude of the mass flux in a vertical section 
$x=83$ at the iteration 1800. The regression were done with the least 
square method. For the exponential model the amplitude was $A=0.0093$, the 
typical depth $d=21.9$ and the residue is of order $10^{-5}$. For the analytical
model, the amplitude is 0.012, the wavelength is$\lambda_R=37.1$ and the residue
is of order $10^{-7}$}
\label{fig:RayleighDepthAmp}
\end{figure}

The exponential decreasing amplitude with depth has been plotted for a vertical 
slice at the time step 1800, at position $x=83$. In fact this time was chosen, 
despite the weak amplitude, because the remaining S waves and other  artefacts 
are clearly separated. The results are displayed in the 
Figure \ref{fig:RayleighDepthAmp}.

The exponential model is simply
\begin{equation}
    j_y=A \exp(-y/d)
\end{equation}

With the exponential regression (fitting with the amplitude $A$ and the typical 
depth $d$), we see that the damping is approximately of the exponential form, 
yet, the amplitude near the surface has a bump. In fact, a rigorous study of the
analytical solution for the Rayleigh equation gives a more complex solution  
\cite{destrade_linear_nodate}: 

\begin{equation}
j_y=-\alpha_S\left[\exp(-k\alpha_Sy)-\frac{2}{1+\alpha_P^2} \exp(-k\alpha_Py)
\right]
\end{equation}
with $\alpha_S\equiv\sqrt{1-v_R^2/v_S^2}$ and $\alpha_P\equiv\sqrt{1-
v_R^2/v_P^2}$.

This analytical model was integrated and fitted with the amplitude and the 
wavelength $\lambda_R$ of the wave (giving $k=2\pi/\lambda_R$). 
It can be observed that the simulation results show a better match with these 
analytical results and the regression even recovers the correct 
wavelength $\lambda_R$ (within 1.3\% error) that can be also measured between 
two maxima of the mass flux.

These results clearly show that the present solid LBM is able to accurately 
capture surface waves, and that the proposed  boundary conditions for rigid 
surfaces, free surfaces and sponge layers are efficient.

\section{Conclusions}

The Lattice Boltzmann method proposed in \cite{jmurthy_lattice_2018} has been 
extended in several ways, by i) using a regular lattice in place of the more 
complex Crystallographic Lattice used in the original version, ii) defining 
boundary conditions for free surface and rigid surfaces and iii) introducing 
efficient sponge layer technique to prevent spurious wave reflection.
These new elements have been supplemented by two theoretical analyses, i.e. the 
Chapman-Enskog expansion to recover the associated macroscopic equations and the 
von-Neumann-type linearized stability analysis.

It is shown that the method the first-order accurate considering the $L_2$ norm 
of the error on the mass flux, and that the method is unconditionnally stable 
for Poisson solids and stable for low-wave-number solution for other values of 
the Poisson ratio $\nu$.

Numerical experiments have demonstrated the accuracy of the present method for 
both bulk and surface waves and, to the knowledge of the authors, it is the 
first time that surface waves are successfully computed using a Lattice 
Boltzmann method for elastic solids.

\section*{Acknowledgments}
The authors warmly acknowledged Dr. G. S. O'Brien for sharing his C codes and 
useful discussions.\\

\section*{Appendix A. Chapman-Enskog Analysis}
The moment $R^{eq}_{\alpha\beta\gamma\kappa}$ for regular lattices is
\begin{align}
\begin{split}
R_{\alpha\beta\gamma\kappa}^{eq}&=\rho b^4 
\Delta^{(4)}_{\alpha\beta\gamma\kappa}+
\frac{1}{2b^4}P^n_{\theta\psi}\left(
\sum_i w_i c_{i\alpha}c_{i\beta}c_{i\gamma}c_{i\kappa}c_{i\theta}c_{i\psi}-
b^2\delta_{\theta\psi}\sum_i w_i c_{i\alpha}c_{i\beta}c_{i\gamma}c_{i\kappa}
\right)\\
&=\rho b^4 \Delta^{(4)}_{\alpha\beta\gamma\kappa}+
\frac{b^2}{2}P^n_{\theta\psi}\left(
\Delta^{(6)}_{\alpha\beta\gamma\kappa\theta\psi} 
-6\delta_{\alpha\beta\gamma\kappa\theta\psi}-
\delta_{\theta\psi}\Delta^{(4)}_{\alpha\beta\gamma\kappa}\right)\\
&=\rho b^4 \Delta^{(4)}_{\alpha\beta\gamma\kappa}+
\frac{b^2}{2}\left(
P^n_{\alpha\psi}\Delta^{(4)}_{\beta\gamma\kappa\psi}+
P^n_{\beta\psi}\Delta^{(4)}_{\alpha\gamma\kappa\psi}+
P^n_{\gamma\psi}\Delta^{(4)}_{\alpha\beta\kappa\psi}+
P^n_{\kappa\psi}\Delta^{(4)}_{\alpha\beta\gamma\psi}-
6 P^n_{\psi\psi}\delta_{\alpha\beta\gamma\kappa\psi}\right)\\
&=\rho b^4 \Delta^{(4)}_{\alpha\beta\gamma\kappa}+
b^2\left(P^n_{\alpha\beta}\delta_{\gamma\kappa}+
P^n_{\alpha\gamma}\delta_{\beta\kappa}+
P^n_{\alpha\kappa}\delta_{\beta\gamma}+
P^n_{\beta\gamma}\delta_{\alpha\kappa}+
P^n_{\beta\kappa}\delta_{\alpha\gamma}+
P^n_{\lambda\kappa}\delta_{\alpha\beta}
-3 P^n_{\psi\psi}\delta_{\alpha\beta\gamma\kappa\psi}\right)
\end{split}
\label{Req}
\end{align}
Taking a spatial derivative $\partial^{(1)}_\kappa$ of the above moment 
gives
\begin{align}
\begin{split}
\partial^{(1)}_\kappa R_{\alpha\beta\gamma\kappa}^{eq}=&
\rho b^4\left(\delta_{\beta\gamma}\partial_\alpha^{(1)}\rho+
\delta_{\alpha\gamma}\partial_\beta^{(1)}\rho+
\delta_{\alpha\beta}\partial_\gamma^{(1)}\rho\right)
+b^2\left(\partial_\alpha^{(1)}P^n_{\beta\gamma}+
\partial_\beta^{(1)}P^n_{\alpha\gamma}+
\partial_\gamma^{(1)}P^n_{\alpha\beta}\right)\\
&+b^2\partial_\kappa^{(1)}\left(
P^n_{\alpha\kappa}\delta_{\beta\gamma}+
P^n_{\beta\kappa}\delta_{\alpha\gamma}+
P^n_{\gamma\kappa}\delta_{\alpha\beta}
-3 P^n_{\psi\psi}\delta_{\alpha\beta\gamma\kappa\psi}\right)\\
=&b^2\left(\partial_\alpha^{(1)}P_{\beta\gamma}+
\partial_\beta^{(1)}P_{\alpha\gamma}+
\partial_\gamma^{(1)}P_{\alpha\beta}+
\partial_\kappa^{(1)}\left(
P^n_{\alpha\kappa}\delta_{\beta\gamma}+
P^n_{\beta\kappa}\delta_{\alpha\gamma}+
P^n_{\gamma\kappa}\delta_{\alpha\beta}
-3 P^n_{\kappa\kappa}\delta_{\alpha\beta\gamma\kappa}\right)\right).
\label{dReq}
\end{split}
\end{align}
The time derivative of $Q_{\alpha\beta\gamma}^{eq}$ can be computed using 
Eq. \eqref{MomentsFeq} and replacing the $\partial_t^{(1)}j_\alpha$ terms with 
$-\partial_\kappa^{(1)}P_{\alpha\kappa}+S_\alpha^{(1)}$ via Eq. 
\eqref{CEmoments1} to be
\begin{align}
\begin{split}
\partial_t^{(1)}Q_{\alpha\beta\gamma}^{eq}=&b^2\partial_t^{(1)}
\left(j_\alpha\delta_{\beta\gamma}+
j_\beta\delta_{\alpha\gamma}+
j_\gamma\delta_{\alpha\beta}\right)\\
=&-b^2\partial_\kappa^{(1)}\left(P_{\alpha\kappa}\delta_{\beta\gamma}+
P_{\beta\kappa}\delta_{\alpha\gamma}+
P_{\alpha\kappa}\delta_{\alpha\beta}\right)\\
&+b^2\left(S^{(1)}_\alpha\delta_{\beta\gamma}+
S^{(1)}_\beta \delta_{\alpha\gamma}+
S^{(1)}_\gamma \delta_{\alpha\beta}
\right)
\end{split}
\label{dQeq}
\end{align}
Using the above relations $Q_{\alpha\beta\gamma}^{(1)}$ can be 
calculated from the final equation of Eq. \eqref{CEmoments1} to be
\begin{align}
\begin{split}
Q_{\alpha\beta\gamma}^{(1)}=&-\tau\left[\partial_t^{(1)}Q_{\alpha\beta\gamma}^{
eq} +\partial^{(1)}_\kappa R_{\alpha\beta\gamma\kappa}^{eq}
-b^2\left(1-\frac{\Delta t}{2\tau}\right)
\left(S^{(1)}_\alpha\delta_{\beta\gamma}+
S^{(1)}_\beta \delta_{\alpha\gamma}+
S^{(1)}_\gamma \delta_{\alpha\beta}
\right)\right]\\
=&-\tau b^2\left[ \partial_\alpha^{(1)}P^n_{\beta\gamma}+
\partial_\beta^{(1)}P^n_{\alpha\gamma}+
\partial_\gamma^{(1)}P^n_{\alpha\beta}
-3\partial_\kappa^{(1)}P_{\kappa\kappa}^n\delta_{\alpha\beta\gamma\kappa}\right]
\\
&-b^2\left[\frac{\Delta t}{2}\left(
S^{(1)}_\alpha\delta_{\beta\gamma}+
S^{(1)}_\beta \delta_{\alpha\gamma}+
S^{(1)}_\gamma \delta_{\alpha\beta}
\right)\right]
\end{split}
\label{Q1}
\end{align}

\section*{Appendix B. Custom Fourier spectral method}
A Fourier spectral method with Crank-Nicholson time integration is used to 
compare bulk waves. We define $(\hat{j_x},\hat{j_y})^T$ as the spatial Fourier 
transform of the mass flux vector $(j_x, j_y)^T$. Let the 
$\hat{J}$ vector be
\begin{equation}
    \hat{J}=
    \begin{pmatrix}
    \hat{j_x}\\
    \hat{j_y}\\
    \partial_t\hat{j_x}\\
    \partial_t\hat{j_y}
    \end{pmatrix}.
    \label{J}
\end{equation}
To simplify the notation, we denote $a^2=(\lambda+2\mu)/\rho_0$,
$b^2=\mu/\rho_0$ and $d^2=(\lambda+\mu)/\rho_0$,  
such that the Navier equation for $j_x$ and $j_y$ in the Fourier domain 
can be written as
\begin{align}
    \begin{split}
        &\partial_{tt}\hat{j_x}=-a^2 k_x^2 \hat{j_x}
        -d^2 k_x k_y \hat{j_y}
        -b^2 k_y^2 \hat{j_x}
        +\frac{1}{\rho_0}\partial_t \hat{F_x}\\
        &\partial_{tt}\hat{j_y}=-b^2 k_x^2 \hat{j_y}
        -d^2 k_x k_y \hat{j_x}
        -a^2 k_y^2 \hat{j_y}
        +\frac{1}{\rho_0}\partial_t \hat{F_y}
    \end{split}
    \label{NavierFourier}
\end{align}
Hence,
\begin{equation}
    \partial_t\hat{J}=
    \begin{pmatrix}
    \partial_t\hat{j_x}\\
    \partial_t\hat{j_y}\\
    -a^2 k_x^2 \hat{j_x}
    -d^2 k_x k_y \hat{j_y}
    -b^2 k_y^2 \hat{j_x}
    +\frac{1}{\rho_0}\partial_t \hat{F_x}\\
    -b^2 k_x^2 \hat{j_y}
    -d^2 k_x k_y \hat{j_x}
    -a^2 k_y^2 \hat{j_y}
    +\frac{1}{\rho_0}\partial_t \hat{F_y}
    \end{pmatrix}
    \label{dJ}
\end{equation}
This equation is of the form $\partial_t\hat{J}=f\left(\hat{J},\hat{F}\right)$. 
The Crank-Nicholson scheme is A-Stable, which allows us to take $\Delta 
t=\Delta 
x$, furthermore, it is a symplectic integrator, which is beneficial for 
equations of the form $\Ddot{X}+\omega^2X=0$ like \eqref{dJ}.

The Crank-Nicholson scheme is
\begin{equation}
    \hat{J}_{n+1}=\hat{J}_n+\frac{\Delta t}{2}\left[
    f\left(\hat{J}_n\right)+
    f\left(\hat{J}_{n+1}\right)\right]
    \label{Crank-Nicholson}
\end{equation}
On simplification, the above equations can be evaluated in matrix form to be
\begin{equation}
    M\hat{J}_{n+1}=N\begin{pmatrix}
    \hat{J}_{n}\\
    \hat{G}_x\\
    \hat{G}_y
    \end{pmatrix},
    \label{MatrixFourier}
\end{equation}
hence,
\begin{equation}
    \hat{J}_{n+1}=M^{-1}N\begin{pmatrix}
    \hat{J}_{n}\\
    \hat{G}_x\\
    \hat{G}_y
    \end{pmatrix}
    \label{Jnext}
\end{equation}
with
\begin{align}
    \begin{split}
        &\hat{G}_x=\frac{1}{\rho_0}\left(\partial_t \hat{F}_{x,n}+
        \partial_t \hat{F}_{x,n+1}\right)\\
        &\hat{G}_y=\frac{1}{\rho_0}\left(\partial_t \hat{F}_{y,n}+
        \partial_t \hat{F}_{y,n+1}\right),
    \end{split}
\end{align}

\begin{equation}
    M=
    \begin{pmatrix}
    1&0&-\frac{\Delta t}{2}&0\\
    0&1&0&-\frac{\Delta t}{2}\\
    \frac{A\Delta t}{2}&\frac{B\Delta t}{2}&1&0\\
    \frac{B\Delta t}{2}&\frac{C\Delta t}{2}&0&1
    \end{pmatrix}
    \label{M}
\end{equation}

\begin{equation}
    N=
    \begin{pmatrix}
    1&0&\frac{\Delta t}{2}&0&0&0\\
    0&1&0&\frac{\Delta t}{2}&0&0\\
    \frac{-A\Delta t}{2}&\frac{-B\Delta t}{2}&1&0&
    \frac{\Delta t}{2}&0\\
    \frac{-B\Delta t}{2}&\frac{-C\Delta t}{2}&0&1&0&
    \frac{\Delta t}{2}
    \end{pmatrix}
    \label{N}
\end{equation}
and,
\begin{align}
    \begin{split}
        &A=a^2 k_x^2+b^2 k_y^2\\
        &B=d^2 k_x k_y\\
        &C=a^2 k_y^2+b^2 k_x^2
    \end{split}
    \label{ABC}
\end{align}

Then the equation \eqref{Jnext} is just a simple matrix relation that can be 
iterated along with the LBM. When needed, the mass flux can be obtained by an 
inverse Fourier transform.

\bibliography{MaximeBib}

\begin{thebibliography}{10}
\expandafter\ifx\csname url\endcsname\relax
  \def\url#1{\texttt{#1}}\fi
\expandafter\ifx\csname urlprefix\endcsname\relax\def\urlprefix{URL }\fi
\expandafter\ifx\csname href\endcsname\relax
  \def\href#1#2{#2} \def\path#1{#1}\fi

\bibitem{kruger_book}
T.~Kru\"uger, H.~Kusumaatmaja, A.~Kuzmin, O.~Shardt, G.~Silva, E.~Viggen, The
  Lattice Boltzmann Method. Principles and Practice, Springer, 2017.

\bibitem{guo_book}
Z.~Guo, C.~Shu, The Lattice Boltzmann Method and its applications in
  engineering, World Scientific, 2013.

\bibitem{Jacob2018b}
J.~Jacob, O.~Malaspinas, P.~Sagaut, A new hybrid recursive regularised
  bhatnagar--gross--krook collision model for lattice boltzmann method-based
  large eddy simulation, Journal of Turbulence 19~(11-12) (2018) 1051--1076.

\bibitem{feng2019hybrid}
Y.~Feng, P.~Boivin, J.~Jacob, P.~Sagaut, Hybrid recursive regularized thermal
  lattice boltzmann model for high subsonic compressible flows, Journal of
  Computational Physics 394 (2019) 82--99.

\bibitem{Guo2020lbm}
S.~Guo, Y.~Feng, J.~Jacob, P.~Sagaut, {An efficient lattice Boltzmann method
  for industrial aerodynamics flows, I: basic model on D3Q19 lattice}, Journal
  of Computational Physics 418 (2020) 109570.

\bibitem{dellar2014}
P.~Dellar, Lattice boltzmann formulation for linear viscoelastic fluids using
  an abstract second stress, SIAM Journal of Scientific Computing 36 (2014).

\bibitem{guangwu2000}
Y.~Guangwu, A lattice boltzmann equation for waves, Journal of Computational
  Physics 161 (2000) 61--69.

\bibitem{jiaung2001}
W.~Jiaung, J.~Ho, C.~Kuo, Lattice boltzmann method for the heat conduction
  problem with phase change, Numerical Heat Transfer, Part B 39 (2001).

\bibitem{zhong2006}
L.~Zhiong, S.~Feng, P.~Dong, S.~Gao, Lattice boltzmann schemes for the
  nonlinear schr\"odinger equation, Physical Review E 74 (2006).

\bibitem{hasanoge2011}
S.~Hasanoge, S.~Succi, S.~Orszag, Lattice boltzmann method for electromagnetic
  wave propagation, EuroPhysics Letters 96 (2011).

\bibitem{liu2014}
Y.~Liu, G.~Yan, A lattice {Boltzmann} {Model} for maxwell's equations, Applied
  Mathematical Modelling 38 (2014).

\bibitem{yin_direct_2016}
X.~Yin, G.~Yan, T.~Li, Direct simulations of the linear elastic displacements
  field based on a lattice {Boltzmann} model: {Direct} simulations of the
  linear elastic displacements field based on a lattice {Boltzmann} model, Int.
  J. Numer. Meth. Engng 107~(3) (2016) 234--251.
\newblock \href {https://doi.org/10.1002/nme.5167}
  {\path{doi:10.1002/nme.5167}}.

\bibitem{obrien_lattice_2012}
G.~S. O'Brien, T.~Nissen-Meyer, C.~J. Bean, A {Lattice} {Boltzmann} {Method}
  for {Elastic} {Wave} {Propagation} in a {Poisson} {Solid}, Bulletin of the
  Seismological Society of America 102~(3) (2012) 1224--1234.
\newblock \href {https://doi.org/10.1785/0120110191}
  {\path{doi:10.1785/0120110191}}.

\bibitem{jmurthy_lattice_2018}
J.~S.~N. J.Murthy, P.~K. Kolluru, V.~Kumaran, S.~Ansumali, Lattice {Boltzmann}
  {Method} for {Wave} {Propagation} in {Elastic} {Solids}, CiCP 23~(4) (2018).
\newblock \href {https://doi.org/10.4208/cicp.OA-2016-0259}
  {\path{doi:10.4208/cicp.OA-2016-0259}}.

\bibitem{xiao2007}
S.~Xiao, A lattice {Boltzmann} method for shock wave propagation in solids,
  Communications in Numerical Methods in Engineering 23 (2007).

\bibitem{namburi_crystallographic_2016}
M.~Namburi, S.~Krithivasan, S.~Ansumali, Crystallographic {Lattice} {Boltzmann}
  {Method}, Sci Rep 6~(1) (2016) 27172.
\newblock \href {https://doi.org/10.1038/srep27172}
  {\path{doi:10.1038/srep27172}}.

\bibitem{kolluru2020lattice}
P.~K. Kolluru, M.~Atif, M.~Namburi, S.~Ansumali, Lattice boltzmann model for
  weakly compressible flows, Physical Review E 101~(1) (2020) 013309.

\bibitem{ispolatov2002}
I.~Ispolatov, M.~Grant, Lattice boltzmann method for viscoelastic fluids,
  Physical Review E 65 (2002).

\bibitem{lallemand2003}
P.~Lallemand, D.~d'Humieres, L.~Luo, R.~Rubinstein, Theory of the lattice
  boltzmann method: Three-dimensional model for the linear viscoelastic fluids,
  Physical Review E 67 (2003).

\bibitem{malaspinas2010}
O.~Malaspinas, N.~F\'etier, M.~Deville, Lattice boltzmann method for the
  simulation of viscoelastic fluid flows, Journal of Non-Newtonian Fluid
  Mechanics 165 (2010).

\bibitem{phillips2011}
T.~Phillips, G.~Roberts, Lattice boltzmann models for non-newtonian flows,
  Journal of Applied Mathematics 76 (2011).

\bibitem{frantziskonis_lattice_2011}
G.~N. Frantziskonis, Lattice {Boltzmann} method for multimode wave propagation
  in viscoelastic media and in elastic solids, Phys. Rev. E 83~(6) (2011)
  066703.
\newblock \href {https://doi.org/10.1103/PhysRevE.83.066703}
  {\path{doi:10.1103/PhysRevE.83.066703}}.

\bibitem{gupta2015}
A.~Gupta, M.~Sbragaglia, A.~Scagliarini, Hybrid lattice boltzmann/finite
  difference simulations of viscoelastic multicomponent flows in confined
  gepmetries, Journal of Computational Physics 291 (2015).

\bibitem{obrien2004}
G.~O'Brien, C.~Bean, A 3d discrete numerical elastic lattice method for seismic
  wave propagation in heterogeneous media with topography, Geophysical Research
  Letters 31 (2004).

\bibitem{obrien2011}
G.~O'Brien, C.~Bean, An irregular lattice method for elastic wave propagation,
  Geophysical Journal International 187 (2011).

\bibitem{obrien2014}
G.~O'Brien, Elastic lattice modelling of seismic waves including a free
  surface, Computers and Geosciences 67 (2014).

\bibitem{obrien2009}
G.~O'Brien, Dispersion analysis and computational efficiency of elastic lattice
  methods for seismic wave propagation, Computers and Geosciences 35 (2009).

\bibitem{vallegarcia2003}
R.~del Valle-Garcia, F.~Sanchez-Sesma, Rayleigh waves modeling using an elastic
  lattice model, Geophysical Research Letters 30 (2003).

\bibitem{xia2017}
M.~Xia, H.~Zhou, Q.~Li, H.~Chen, Y.~Wang, S.~Wang, A general 3d lattice spring
  model for modeling elastic waves, Bulletin of the Seismological Society of
  America 107 (2017).

\bibitem{polyzos2012}
D.~Polyzos, D.~Fotiadis, Derivation of mindlin's first and second strain
  gradient elastic theory via simple lattice and continuum models,
  International Journal of Solids and Structures 49 (2012).

\bibitem{godunov2003}
S.~Godunov, E.~Romenskii, Elements of continuum mechanics and conservation
  laws, Springer, 2003.

\bibitem{chen_fundamental_2008}
H.~Chen, X.~Shan, Fundamental conditions for {N}-th-order accurate lattice
  {Boltzmann} models, Physica D: Nonlinear Phenomena 237~(14-17) (2008)
  2003--2008.
\newblock \href {https://doi.org/10.1016/j.physd.2007.11.010}
  {\path{doi:10.1016/j.physd.2007.11.010}}.

\bibitem{sengupta2007}
T.~Sengupta, A.~Dipankar, P.~Sagaut, Error dynamics: Beyond von {N}eumann
  analysis, Journal of Computational Physics 226 (2007).

\bibitem{wissocq_extended_2019}
G.~Wissocq, P.~Sagaut, J.-F. Boussuge, An extended spectral analysis of the
  lattice {Boltzmann} method: modal interactions and stability issues, Journal
  of Computational Physics 380 (2019) 311--333.
\newblock \href {https://doi.org/10.1016/j.jcp.2018.12.015}
  {\path{doi:10.1016/j.jcp.2018.12.015}}.

\bibitem{xia_modelling_2017}
M.~Xia, S.~Wang, H.~Zhou, X.~Shan, H.~Chen, Q.~Li, Q.~Zhang, Modelling
  viscoacoustic wave propagation with the lattice {Boltzmann} method, Sci Rep
  7~(1) (2017) 10169.
\newblock \href {https://doi.org/10.1038/s41598-017-10833-w}
  {\path{doi:10.1038/s41598-017-10833-w}}.

\bibitem{aki_quantitative_2009}
K.~Aki, P.~G. Richards, Quantitative seismology, 2nd Edition, Univ. Science
  Books, Sausalito, Calif, 2009, oCLC: 845610339.

\bibitem{destrade_linear_nodate}
M.~Destrade, G.~Saccomandi, Linear {Elastodynamics} and {Waves}  36.

\end{thebibliography}

\end{document}